\documentclass[11pt,draftclsnofoot,onecolumn]{IEEEtran}
\usepackage{amsfonts}
\usepackage[pdftex]{graphicx}
\usepackage{epsfig,latexsym}
\usepackage{float}
\usepackage{indentfirst}
\usepackage{color}

\usepackage{amssymb}
\usepackage{amsmath}
\usepackage{times}
\usepackage{subfigure}
\usepackage{psfrag}
\usepackage{cite}
\usepackage{booktabs}
\usepackage{multirow}
\usepackage{verbatim}
\usepackage{diagbox}
\usepackage{bbm}

\newtheorem{theorem}{$\mathbf{Theorem}$}
\newtheorem{lemma}{$\mathbf{Lemma}$}

\newtheorem{corollary}{$\mathbf{Corollary}$}

\newtheorem{remark}{$\mathbf{Remark}$}

\newcounter{tempEquationCounter} 
\newcounter{thisEquationNumber}

\begin{document}
	
	\title{Understanding Deep MIMO Detection}
	
	\author{Qiang~Hu,~Feifei~Gao,~\IEEEmembership{Fellow,~IEEE},~Hao Zhang,~Geoffrey~Ye~Li,~\IEEEmembership{Fellow,~IEEE},~and~Zongben~Xu% <-this % stops a spac}
		
		\thanks{Q. Hu and H. Zhang are with the Department
			of Electrical Engineering, Tsinghua University, Beijing 100084, P. R. China (e-mail: huq16@mails.tsinghua.edu.cn; haozhang@mail.tsinghua.edu.cn).}% <-this % stops a space
		\thanks{F. Gao is with the Institute for Artificial Intelligence, Tsinghua University (THUAI), 
			State Key Lab of Intelligent Technologies and Systems, Tsinghua University, 
			Beijing National Research Center for Information Science and Technology (BNRist), 
			Department of Automation, Tsinghua University,
			Beijing 100084, P. R. China (e-mail: feifeigao@ieee.org).}
		\thanks{G. Y. Li is with the Department of Electrical and Computer Engineering, Imperial College London, London, UK (e-mail: Geoffrey.Li@imperial.ac.uk).}
		\thanks{Z. Xu is with the School of Mathematics and Statistics, Xi'an Jiaotong University, Xi'an 710049, P. R. China (e-mail: zbxu@mail.xjtu.edu.cn).}}
	\maketitle
	\vspace{-4mm}
	\begin{abstract}
		Incorporating deep learning (DL) into multiple-input multiple-output (MIMO) detection has been deemed as a promising technique for future wireless communications. However, most DL-based detection algorithms are lack of theoretical interpretation on internal mechanisms and could not provide general guidance on network design. In this paper, we analyze the performance of DL-based MIMO detection to better understand its strengths and weaknesses. We investigate two different architectures: a data-driven DL detector with a neural network activated by rectifier linear unit (ReLU) function and a model-driven DL detector from unfolding a traditional iterative detection algorithm. We demonstrate that data-driven DL detector asymptotically approaches to the maximum a posterior (MAP) detector in various scenarios but requires enough training samples to converge in time-varying channels. On the other hand, the model-driven DL detector utilizes model expert knowledge to alleviate the impact of channels and establish a relatively reliable detection method with a small set of training data. Due to its model specific property, the performance of model-driven DL detector is largely determined by the underlying iterative detection algorithm, which is usually suboptimal compared to the MAP detector. Simulation results confirm our analytical results and demonstrate the effectiveness of DL-based MIMO detection for both linear and nonlinear signal systems.
	\end{abstract}
	
	
	\IEEEpeerreviewmaketitle
	%\vspace{50mm}
	\section{Introduction}
	\label{introduction}
	Multiple-input multiple-output (MIMO) technology is vital for modern wireless communication systems to support explosively growing throughput requirement ~\cite{marzetta2010,larsson2014,bolei2018}. In general, the maximum a \textit{posterior} (MAP) detector delivers the optimal detection performance but has an exponential computational complexity, which is infeasible for large-sized MIMO systems~\cite{1998Multiuser}. Alternatively, suboptimal detection algorithms are implemented to achieve a better tradeoff between accuracy and complexity. The linear detectors, such as matched filter (MF), zero-forcing (ZF), and linear minimum mean-squared error (LMMSE), are with low complexity but exhibit poor performance compared to the MAP detector. On the other hand, iterative detection algorithms, e.g., approximate message passing (AMP)~\cite{donoho2009message,bayati2011}, sphere decoding (SD)~\cite{renqiu2006}, soft interference cancellation (SIC)~\cite{xiaodong1999,choi2000}, can achieve good performance with moderate complexity under some practical scenarios. All these detectors are model-specific and require complete knowledge of channel state information (CSI), which are prone to error propagation and suffer from serious performance deterioration if the system model mismatches the real transmission model or if the imperfect CSI is presented.

	Over the last decade, deep learning (DL) has made profound technical revolution to many areas, such as computer vision~\cite{krizhevsky2012} and speech recognition~\cite{hinton2012}. Inspired by these successes, DL has been  applied to the design of communication systems recently, including physical layer processing~\cite{qin2019deep} (channel estimation~\cite{yang2019MIMO,ye2020deep} and symbol detection~\cite{o2017introduction}), and resource management~\cite{ye2019deep,liang2019deep}, etc. Among all DL-based applications, MIMO detection is one of the most crucial and fundamental issues. DL-based detectors could learn to map the received signals into the transmitted symbols from training data and achieve better performance than the traditional detection algorithms~\cite{ye2018power}.
	
	Generally, DL-based detectors can be divided into two categories: data-driven DL detectors based on deep neural networks (DNNs) and model-driven DL detectors from unfolding iterative detection algorithms. Data-driven DL detectors use DNN architectures to implement symbol detection~\cite{farsad2018,nir2021}. These DNN embedded architectures are model independent and can recover transmitted symbols in various scenarios with high precision if properly trained. However, such properties come at the price of a large number of trainable parameters and training samples. On the other hand, model-driven DL detectors are designed from the traditional iterative detection algorithms, where each layer of the network represents a single iteration with some trainable variables added~\cite{he2019model,Samuel2019Learning,hedetec2020}. The resulting detectors tend to have better performance and faster convergence compared to original iterative detection algorithms~\cite{hedetec2020}. However, current model-driven DL detectors are established on the premise that channel model is linear and CSI is available, limiting their application in complicated environments.
	
	Despite their great success, data-driven DL detectors are considered as black boxes for signal reception and only experimental evaluation is available to demonstrate their performance. It is desired to understand internal mechanism of DL-based MIMO detection and provide a general design guidance. In fact, there exits a lot of literature on analyzing internal mechanisms of DNNs. The pioneering works in~\cite{cybenko1989approximation,hornik1989multilayer} have proved that any continuous function on a compact set can approximated with any precision by a DNN with sigmoid activation function. Recently, it has been proved in~\cite{montufar2014number,arora2018understanding} that DNNs with rectified linear units (ReLU DNNs) can also approximate to a large family of functions. Furthermore, DL-based channel estimation has been proved to converges to the minimum mean-squared error (MMSE) estimator as the size of training set increases in~\cite{qiang2020channel}. However, MIMO detection is a classification problem and the analysis of DL-based channel estimation in~\cite{qiang2020channel} cannot be directly generalized to DL-based MIMO detection. To the best of the authors' knowledge, there is no analytical interpretation to the advantages and disadvantages of DL-based MIMO detection.
	
	In this paper, we analyze the performance of DL-based MIMO detection including the data-driven and the model-driven DL detectors. Our contributions are listed as follows.
	\begin{itemize}
		\item We prove that the data-driven DL detector with ReLU DNN can well approximate the MAP detector under sufficiently large training set in MIMO systems. The rate of convergence of the DL detector to the MAP detector scales at least polynomially fast with the size of training samples.  
		\item We show that the data-driven DL detector requires no CSI to approach the MAP detector for time-invariant channels and is robust to CSI uncertainty. For time-varying channels, the data-driven DL detector requires perfect CSI to converge to the MAP detector and is sensitive to CSI uncertainty.
		\item We prove that the model-driven DL detector may asymptotically approach to the optimal one that minimizes the mean-squared error (MSE) or the expectation of Kullback-Leibler (KL) divergence as the size of training set increases if the original iterative detection algorithm is properly designed. In general, the model-driven DL detector requires much less training data but has lower detection accuracy than the  data-driven DL detector.
	\end{itemize}

	The rest of this paper is organized as follows. The system model and the traditional MIMO detection algorithms are introduced in Section~\ref{preliminary}. The performance analysis of the data-driven and the model-driven DL detectors is presented in Sections~\ref{linear-sec} and~\ref{model-driven}, respectively. Simulation results are provided in Section~\ref{simulation} followed by the conclusions in Section~\ref{conclusion-sec}.
	
	\textit{Notations}: We use lowercase letters and capital letters in boldface to denote vectors and matrices, respectively. The positive integer set, natural number set, real number set, and complex number set are denoted by $\mathbb{N}$, $\mathbb{Z}$, $\mathbb{R}$, and $\mathbb{C}$, respectively. The real and imaginary parts of a complex matrix or vector are defined by $\Re(\cdot)$ and $\Im(\cdot)$, respectively. $\mathbf{I}_{M}$ denotes the $M\times M$ identity matrix. Notations $(\cdot)^{T}$ and $(\cdot)^{H}$ represent the transpose and Hermitian of a matrix or a vector, respectively. $\mathbb{E}\{\cdot\}$ denotes the expectation, $\mathrm{tr}\{\cdot\}$ denotes the trace of a matrix, and $\mathrm{vec}(\cdot)$ denotes the vectorization of a matrix. The cardinality of a set is denoted by $|\cdot|$. Notations $\|\cdot\|_{2}$  and $\|\cdot\|_{\infty}$ represent the $2$-norm and supremum-norm of a vector or a matrix, respectively. Notation $\lceil\cdot\rceil$ represents the ceiling of a real number. Notation $\mathbf{1}_{\mathcal{A}}(x)$ is an indicator function of set $\mathcal{A}$, where $\mathbf{1}_{\mathcal{A}}(x)=1$ if $x\in\mathcal{A}$ and $\mathbf{1}_{\mathcal{A}}(x)=0$ if $x\notin\mathcal{A}$.
	
	\section{Traditional MIMO Detection}\label{preliminary}
	In this section, we first introduce a MIMO communication system and then present some traditional MIMO detection algorithms. 
	\subsection{System Model}
	Let us consider a standard linear MIMO system with $d_{t}$ transmit and $d_{r}$ receive antennas. The $d_{r}\times 1$ received signal vector at the BS is
	\begin{align}\label{signal_model}
		\bar{\mathbf{x}}=\bar{\mathbf{H}}\bar{\mathbf{s}}+\bar{\mathbf{n}},
	\end{align}
	where $\bar{\mathbf{H}}\in\mathbb{C}^{d_{r}\times d_{t}}$ is the channel matrix, $\bar{\mathbf{s}}\in\bar{\mathbb{S}}^{d_{t}}$ is a transmitted symbol vector of mutually independent elements drawn from a discrete constellation $\bar{\mathbb{S}}$, and $\bar{\mathbf{n}}\in\mathbb{C}^{d_{t}}$ is an independent zero-mean Gaussian noise vector with element-wise variance $\sigma_{n}^{2}$.
	
	To avoid handling complex values in MIMO detection, we re-parameterize~\eqref{signal_model} into a real-valued signal model,
	\begin{align}\label{signal_model2}
		\mathbf{x}=\mathbf{H}\mathbf{s}+\mathbf{n},
	\end{align}
	where
	\begin{align}\nonumber
		\mathbf{x}=\begin{bmatrix}\Re(\bar{\mathbf{x}})\\\Im(\bar{\mathbf{x}})\end{bmatrix},		\mathbf{s}=\begin{bmatrix}\Re(\bar{\mathbf{s}})\\\Im(\bar{\mathbf{s}})\end{bmatrix},
		\mathbf{n}=\begin{bmatrix}\Re(\bar{\mathbf{n}})\\\Im(\bar{\mathbf{n}})\end{bmatrix},
	\end{align}
	and
	\begin{align}\label{real_part}
		\mathbf{H}=\begin{bmatrix}\Re(\bar{\mathbf{H}})&-\Im(\bar{\mathbf{H}})\\\Im(\bar{\mathbf{H}})&\Re(\bar{\mathbf{H}})\end{bmatrix}.
	\end{align}
	
	Denote $\mathbb{S}=\Re(\bar{\mathbb{S}})$ as the real part of $\bar{\mathbb{S}}$ and assume that $\Re(\bar{\mathbb{S}})=\Im(\bar{\mathbb{S}})$. Then, we have $\mathbf{s}\in\mathbb{S}^{2d_{t}}$ in~\eqref{signal_model2}.

	\subsection{Traditional MIMO Detection}\label{traditional}
	The following examples are some traditional MIMO detection algorithms.
	
	\subsubsection{MAP Detector} Let $p_{o}(\mathbf{s}|\mathbf{x})$ be the \textit{posterior} probability of $\mathbf{s}$ given $\mathbf{x}$. The MAP detector is optimal in terms of minimizing the error probability of symbol detection given $\mathbf{x}$~\cite{kay1993fundamentals}, i.e.,
	\begin{align}\label{map_detec}
		\mathbf{s}_{\mathrm{MAP}}=\arg\max_{\mathbf{s}\in\mathbb{S}^{2d_{t}}}p_{o}(\mathbf{s}|\mathbf{x}).
	\end{align}
	
	The MAP detector in~\eqref{map_detec} is equivalent to the maximum likelihood (ML) detector when the transmitted symbols are with equal probability. However, there are two reasons that prevent the MAP detector from practical applications: 1) the optimization in~\eqref{map_detec} requires an exhausted search of $|\mathbb{S}|^{2d_{t}}$ different possible input combinations and is computationally infeasible especially when $d_{t}$ is large; 2) accurate knowledge of $p_{o}(\mathbf{s}|\mathbf{x})$ is required to implement the MAP detector, which is sometimes very hard. 
	
	\subsubsection{ZF Detector}
	A common strategy for decoding $\mathbf{s}$ with affordable computational complexity is to utilize the ZF detector~\cite{1998Multiuser},
	\begin{align}\label{zf_detec}
		\mathbf{s}_{\mathrm{ZF}} = (\mathbf{H}^{T}\mathbf{H})^{-1}\mathbf{H}^{T}\mathbf{x}.
	\end{align}
	
	The ZF detector involves only simple matrix computations and is easy to implement in practice. However, such simplicity comes at the cost of low accuracy. The performance of the ZF detector degrades significantly when the MIMO system is nonlinear or when only imperfect CSI is available. 
	\subsubsection{Iterative Detector}
	To better balance computational complexity and accuracy, iterative detector is used for MIMO detection. Any iterative detection algorithm, such as AMP~\cite{bayati2011} and SIC~\cite{choi2000} algorithms, can be expressed in a cascaded function as
	\begin{align}\label{iterative}
		\mathbf{f}_{\mathrm{iter}}(\mathbf{H},\mathbf{x})=\tilde{\mathcal{A}}_{l_{u}}\circ\tilde{\mathcal{A}}_{l_{u}-1}\circ\cdots\circ\tilde{\mathcal{A}}_{1}(\mathbf{x},\mathbf{H}),
	\end{align}
	where $l_{u}\in\mathbb{N}$ is the iteration number and $\tilde{\mathcal{A}_{i}}$ is the $i$-th iteration for $i\in\{1,\ldots,l_{u}\}$. Generally, each iteration in~\eqref{iterative}
	is with low complexity and increasing $l_{u}$ can continuously improve the detection accuracy until the iterative detector converges. For most of the traditional iterative detectors, $\tilde{\mathcal{A}}_{i}$ keeps the same for different $i$'s and their performance is suboptimal compared to the MAP detector, which has much room for improvement.

	\section{Data-driven DL Detector}\label{linear-sec}
	
	With powerful learning ability, the data-driven DL detector can establish a stable and precise model to achieve the performance comparable with the MAP detector and has been used for MIMO detection. In this section, the performance of the data-driven DL detector is analyzed from a theoretical perspective via statistical learning theory. 
	\subsection{Basic Setting of Data-driven DL Detector without CSI}\label{basic}

	Let us consider a data-driven DL detector, $\mathcal{D}$, with a fully-connected ReLU DNN, where only the received signal $\mathbf{x}$ is available.\footnote{The fully-connected ReLU DNN is the basis for most of current state-of-the-art DNNs~\cite{szegedy2015going,he2016deep}. In this respect, we choose the fully-connected ReLU DNN as an example to analyze the performance of the data-driven DL detector, which can be easily extended to other more advanced network structures.} To consider MIMO detection for general systems, we extend the linear model in~\eqref{signal_model} to the following statistical models
	\begin{align}\label{nonlinear1}
		\mathbf{x}=\mathbf{f}_{\mathrm{nlr}}(\mathbf{H}\mathbf{s}+\mathbf{n}),
	\end{align}
	and
	\begin{align}\label{nonlinear2}
		\mathbf{x}=\mathbf{H}\mathbf{f}_{\mathrm{nlt}}(\mathbf{s})+\mathbf{n},
	\end{align}
	where $\mathbf{f}_{\mathrm{nlr}}(\cdot):\mathbb{R}^{2d_{r}}\rightarrow\mathbb{R}^{2d_{r}}$ and $\mathbf{f}_{\mathrm{nlt}}(\cdot):\mathbb{R}^{2d_{t}}\rightarrow\mathbb{R}^{2d_{t}}$ are the unknown distortions imposed on the received and transmitted signals, respectively, e.g., imperfect power amplifier (PA) at the transmitters~\cite{costa1999impact,costa2002m} or quantization error of analog-to-digital converter (ADC) at the receivers~\cite{XULIANG2019}.
	
	A major concern for $\mathcal{D}$ is that the constellations in communication systems are generally not taken as the targets of DNNs. To comply with standard processing in DL methods, we need to re-parameterize $\mathbf{s}$ using one-hot mapping. Let $\mathbf{s}_{i}\in\mathbb{S}^{2}$ be the $2$-dimensional vector of real-valued symbols transmitted at the $i$-th antenna. Stacking all the symbols at transmitted antenna, $\mathbf{s}$ can be expressed as
	\begin{align}\label{one-hot}
		\mathbf{s}=(s_{j_{1}},s_{j_{2}},\ldots,s_{j_{d_{t}}},s_{j_{d_{t}+1}},\ldots,s_{j_{2d_{t}}}), 
	\end{align}
	where $\{(j_{1},\ldots,j_{2d_{t}})\in\mathbb{N}^{2d_{t}}:0\leq j_{i}, j_{d_{t}+i}\leq (|\mathbb{S}|-1)\ \forall i=1,\ldots,d_{t}\}$ and $\mathbf{s}_{i}=(s_{j_{i}},s_{j_{d_{t}+i}})$ for $s_{j_{i}}\in\mathbb{S}$ and $s_{j_{d_{t}+i}}\in\mathbb{S}$.
	For notation convenience, we associate a unit vector $\mathbf{u}\in\mathbb{R}^{|\mathbb{S}|^{2d_{t}}}$ with each $\mathbf{s}\in\mathbb{S}^{2d_{t}}$ and the index of nonzero element of $\mathbf{u}$ can be derived from $\big[\sum_{i=0}^{d_{t}-1}(j_{d_{t}+i}|S|+j_{i})|\mathbb{S}|^{2i}+1\big]$.
	
	In this way, $\mathbf{u}$ is a bijective transformation of $\mathbf{s}$ and we have $p_{o}(\mathbf{s}|\mathbf{x})=p_{o}(\mathbf{u}|\mathbf{x})$. Hence, one can also set $\mathbf{u}$ as the target for MIMO detection. The input-output sample set of $\mathcal{D}$ is then defined by
	\begin{align}
		\mathcal{Z}=\{(\mathbf{x}_{m}, \mathbf{u}_{m})\,|\,\mathbf{x}\in\mathbb{R}^{2d_{r}},\mathbf{u}\in\mathbb{R}^{|\mathbb{S}|^{2d_{t}}}, m=1,\ldots,|\mathcal{Z}|\}
	\end{align}
and samples in $\mathcal{Z}$ are independent and identically distributed (i.i.d.).
	
	The DNN of $\mathcal{D}$ consists of input and output layers, $l\in\mathbb{N}$ hidden layers, and neuron assignment $\mathbf{d}=(d_{0},d_{1},\ldots,d_{l},d_{l+1})\in\mathbb{N}^{l+1}$ with $d_{0}=2d_{r}$ and $d_{l+1}=|\mathbb{S}|^{2d_{t}}$. The depth, width, and size of $\mathcal{D}$ are defined by $l$, $\max\{d_{1},\ldots,d_{l}\}$, and $d_{u}=\sum_{i}^{l}d_{i}$, respectively.
	
	Let
	\begin{align}\label{parameter}
		\Theta=\{\boldsymbol{\theta}=(\mathrm{vec}(\mathbf{W}_{0}),\mathbf{b}_{0},\ldots,\mathrm{vec}(\mathbf{W}_{l}),\mathbf{b}_{l})\in\mathbb{R}^{d_{s}}\}
	\end{align}
	be the set of all parameters of $\mathcal{D}$, where $d_{s}=\sum_{i=0}^{l}d_{i+1}\times (d_{i}+1)$, $\mathbf{W}_{i}\in\mathbb{R}^{d_{i+1}\times d_{i}}$ is the weight matrix connecting the $i$-th layer to the $(i+1)$-th layer, and $\mathbf{b}_{i}\in\mathbb{R}^{d_{i+1}}$ is the bias vector of the $(i+1)$-th layer for $i\in\{0,\ldots,l\}$. 
	
	For a fixed  $\mathbf{d}$, 
	\begin{align}\label{network_output}
		\mathbf{p}_{\boldsymbol{\theta}}(\mathbf{x})=\psi_{d_{l+1}}\circ\mathcal{A}_{l}\circ\varphi_{d_{l}}\circ\mathcal{A}_{l-1}\circ\varphi_{d_{l-1}}\circ\cdots\circ\varphi_{d_{1}}\circ\mathcal{A}_{0}(\mathbf{x})
	\end{align} 
	is the underlying function of $\mathcal{D}$, where $\circ$ denotes the function composition, $\psi_{d_{l+1}}:\mathbb{R}^{d_{l+1}}\rightarrow\mathbb{R}^{d_{l+1}}$ is the entry-wise softmax function, $\mathcal{A}_{i}:\mathbb{R}^{d_{i}}\rightarrow\mathbb{R}^{d_{i+1}}$ is the affine transformation with weight $\mathbf{W}_{i}$ and bias $\mathbf{b}_{i}$, and $\varphi_{d_{i}}:\mathbb{R}^{d_{i}}\rightarrow\mathbb{R}^{d_{i}}$ is the entry-wise ReLU activation function for $i\in\{0,\ldots,l\}$. 
	
	Denote $\tilde{\mathbf{x}}_{i}=[x_{i,1},\ldots,x_{i,d_{i}}]^{T}$ as the output of the $i$-th layer with $\tilde{\mathbf{x}}_{0}=\mathbf{x}$ and $\mathcal{A}_{i-1}(\tilde{\mathbf{x}}_{i-1})=(x_{\mathcal{A}_{i-1},1},\ldots,x_{\mathcal{A}_{i-1},d_{i}})\in\mathbb{R}^{d_{i}}$. The ReLU activation function and softmax function can be expressed as
	\begin{align}\label{relu}
		\varphi_{d_{i}}(\mathcal{A}_{i-1}(\tilde{\mathbf{x}}_{i-1}))=(\max\{0,x_{\mathcal{A}_{i-1},1}\},\ldots,\max\{0,x_{\mathcal{A}_{i-1},d_{i}}\}),
	\end{align} and
	\begin{align}\label{softmax}
		\psi_{d_{l+1}}(\mathcal{A}_{l}(\tilde{\mathbf{x}}_{l}))=\bigg(\frac{\exp(x_{\mathcal{A},1})}{\sum_{j=1}^{d_{l+1}}\exp(x_{\mathcal{A},j})},\ldots,\frac{\exp(x_{\mathcal{A},d_{l+1}})}{\sum_{j=1}^{d_{l+1}}\exp(x_{\mathcal{A},j})}\bigg),
	\end{align} respectively.
	
	From~\eqref{relu}, the neurons in $\mathcal{D}$ have only two states: zero output or replicating input. All possible states of neurons in $\mathcal{D}$ can be represented by a set $\mathcal{K}\subseteq\{0,1\}^{d_{u}}$ when $\boldsymbol{\theta}$ is fixed. Each element in $\mathcal{K}$ is a $d_{u}$-dimensional vector with its entries being either $0$ or $1$. Similar to~\cite{montufar2014number}, the input space of $\mathcal{D}$ is partitioned into linear regions according to different states. Denote $\mathcal{X}$ and $\mathcal{X}_{k}$ as the input space and the input region corresponding to the $k$-th state, respectively. It is obvious that
	\begin{align}\label{partition}
		&\mathcal{X}_{k}\subseteq\mathcal{X},\ k=1,\ldots,K=|\mathcal{K}|,\nonumber\\
		&\mathcal{X}=\cup_{k=1}^{K}\mathcal{X}_{k}.
	\end{align}

	For $\mathbf{x}\in\mathcal{X}_{k}$, we know $\mathcal{A}_{i}(\tilde{\mathbf{x}}_{i})$ in~\eqref{network_output} satisfies
	\begin{align}\label{hyperplane1}
		\mathcal{A}_{i}(\tilde{\mathbf{x}}_{i})=\left\{\begin{array}{lcl}\mathbf{W}_{0}\mathbf{x}+\mathbf{b}_{0}, & & i=0,\\
			\tilde{\mathbf{W}}_{i}\mathcal{A}_{i-1}(\tilde{\mathbf{x}}_{i-1})+\mathbf{b}_{i},& &i\geq 1,\end{array}\right.
	\end{align}
	where $\tilde{\mathbf{W}}_{i}=\mathbf{W}_{i}\mathbf{\Lambda}_{i}$ and $\mathbf{\Lambda}_{i}$ is an $\mathbb{R}^{d_{i}\times d_{i}}$ diagonal matrix whose diagonal element is either $0$ or $1$. Moreover, $\mathbf{\Lambda}_{0}=\mathbf{I}_{d_{0}}$ and  $\tilde{\mathbf{W}}_{0}=\mathbf{W}_{0}\mathbf{\Lambda}_{0}$. The diagonal elements of  $\mathbf{\Lambda}_{i}$ correspond to the states of neurons at the $i$-th layer. By expanding $\mathcal{A}_{i}(\tilde{\mathbf{x}}_{i})$ recursively, we further obtain
	\begin{align}\label{hyperplane}
		\mathcal{A}_{i}(\tilde{\mathbf{x}}_{i})&=\prod_{j=0}^{i}\tilde{\mathbf{W}}_{j}\mathbf{x}+\sum_{j=0}^{i-1}\bigg(\prod_{p=0}^{j}\tilde{\mathbf{W}}_{i-p}\bigg)\mathbf{b}_{i-1-j}+\mathbf{b}_{i}=\hat{\mathbf{W}}_{i}\mathbf{x}+\hat{\mathbf{b}}_{i},
	\end{align}
	where $\hat{\mathbf{W}}_{i}=\prod_{j=0}^{i}\tilde{\mathbf{W}}_{j}$ and
	$\hat{\mathbf{b}}_{i}=\sum_{j=0}^{i-1}\left(\prod_{p=0}^{j}\tilde{\mathbf{W}}_{i-p}\right)\mathbf{b}_{i-1-j}+\mathbf{b}_{i}$. Let $\mathbf{f}_{\boldsymbol{\theta}}\left(\mathbf{x}\right)=\mathcal{A}_{l}(\tilde{\mathbf{x}}_{l})$ be the input at the last layer. Using~\eqref{hyperplane}, we could derive the explicit form of  $\mathbf{f}_{\boldsymbol{\theta}}(\mathbf{x})$ as
	\begin{align}\label{network} \mathbf{f}_{\boldsymbol{\theta}}(\mathbf{x})=\mathbf{W}_{\mathcal{X}_{k}}\mathbf{x}+\mathbf{b}_{\mathcal{X}_{k}}
	\end{align}
	for any $\mathbf{x}\in\mathcal{X}_{k}$, where $\mathbf{W}_{\mathcal{X}_{k}}=\hat{\mathbf{W}}_{l}$ and $\mathbf{b}_{\mathcal{X}_{k}}=\hat{\mathbf{b}}_{l}$. From~\eqref{network}, $\mathbf{f}_{\boldsymbol{\theta}}(\mathbf{x})$ turns into an affine function for any $\mathbf{x}\in\mathcal{X}_{k}$ and a piecewise linear function for any $\mathbf{x}\in\mathcal{X}$. 
	
	Applying the softmax function to $\mathbf{f}_{\boldsymbol{\theta}}(\mathbf{x})$ yields
	\begin{align}\label{network2} \mathbf{p}_{\boldsymbol{\theta}}(\mathbf{x})=\psi_{d_{l+1}}(\mathbf{f}_{\boldsymbol{\theta}}(\mathbf{x}))=\psi_{d_{l+1}}(\mathbf{W}_{\mathcal{X}_{k}}\mathbf{x}+\mathbf{b}_{\mathcal{X}_{k}}).
	\end{align}
	Each entry of $\mathbf{p}_{\boldsymbol{\theta}}\left(\mathbf{x}\right)$ is restricted in the range $[0,1]$ and the sum of all these entries is equal to $1$, as shown in~\eqref{softmax}. Since $\mathbf{u}$ is the target for detection, $\mathbf{p}_{\boldsymbol{\theta}}\left(\mathbf{x}\right)$ can be regarded as a set of estimated \textit{posterior} probabilities for all possible $\mathbf{u}$ given $\mathbf{x}$. The goal of the data-driven DL detector is to approximate $p_{o}(\mathbf{u}|\mathbf{x})$ by optimizing $\boldsymbol{\theta}$ within some feasible set.

	\subsection{Performance Analysis without CSI}\label{wCSI}
	Unlike model-based detectors in~\eqref{map_detec} and~\eqref{zf_detec}, it is difficult to derive an explicit analytical form of $\mathbf{p}_{\boldsymbol{\theta}}(\mathbf{x})$ and the performance analysis of the data-driven DL detector is not straightforward. 
	
	Let $\tilde{\mathbf{u}}_{i}\in\mathbb{R}^{d_{l+1}}$ denote the unit vector whose $i$-th entry is nonzero. Note that $(\tilde{\mathbf{u}}_{1},\ldots,\tilde{\mathbf{u}}_{d_{l+1}})$ is an i.i.d. multinomial random variable with probability $((p(\tilde{u}_{1}),\ldots,p(\tilde{u}_{d_{l+1}}))$. 
	
	Denote
	\begin{align}\label{posterior}
		p_{\boldsymbol{\theta}}(\mathbf{u}|\mathbf{x})=\prod_{i=1}^{d_{l+1}}p_{\boldsymbol{\theta},i}(\mathbf{x})^{\mathbf{1}_{\tilde{\mathbf{u}}_{i}}(\mathbf{u})}
	\end{align}
	as the estimated \textit{posterior} probability of the data-driven DL detector, where $p_{\boldsymbol{\theta},i}(\mathbf{x})$ is the $i$-th entry of $\mathbf{p}_{\boldsymbol{\theta}}(\mathbf{x})$.
	
	To find $\boldsymbol{\theta}$ that leads to the optimal data-driven DL detector, we need a loss function to measure the distance between $p_{\boldsymbol{\theta}}(\mathbf{u}|\mathbf{x})$ and $p_{o}(\mathbf{u}|\mathbf{x})$. Typically, the KL divergence of $p_{\boldsymbol{\theta}}(\mathbf{u}|\mathbf{x})$ and $p_{o}(\mathbf{u}|\mathbf{x})$ is adopted, which is defined as
	\begin{align}\label{kl-info}
		D_{\mathrm{KL}}(p_{o},p_{\boldsymbol{\theta}})=\mathbb{E}\Big\{\ln\frac{p_{o}(\mathbf{u}|\mathbf{x})}{p_{\boldsymbol{\theta}}(\mathbf{u}|\mathbf{x})}\Big\}.
	\end{align}
	Note that $D_{\mathrm{KL}}(p_{o},p_{\boldsymbol{\theta}})$ is non-negative and is equal to zero if and only if $p_{\boldsymbol{\theta}}(\mathbf{u}|\mathbf{x})=p_{o}(\mathbf{u}|\mathbf{x})$~\cite{cover1999elements}. Though the KL information is not a distance function, its convergence often implies the same trend in other metrics. 
	
	A convenient distance metric is the Hellinger metric:
	\begin{align}
		\gamma(p_{o},p_{\boldsymbol{\theta}})=\Big(\frac{1}{2}\mathbb{E}\big\{(p_{o}(\mathbf{u}|\mathbf{x})^{1/2}-p_{\boldsymbol{\theta}}(\mathbf{u}|\mathbf{x})^{1/2})^{2}\big\}\Big)^{1/2}
	\end{align}
	and the following inequality
	\begin{align}\label{helling}
		\gamma^{2}(p_{o},p_{\boldsymbol{\theta}})\leq \frac{1}{2}D_{\mathrm{KL}}(p_{o},p_{\boldsymbol{\theta}})
	\end{align}
	holds between $D_{\mathrm{KL}}(p_{o},p_{\boldsymbol{\theta}})$ and $\gamma(p_{o},p_{\boldsymbol{\theta}})$~\cite[Lemma 1.3]{vandeer2006}. From~\eqref{helling}, decreasing the KL information reduces the distance between $p_{\boldsymbol{\theta}}(\mathbf{u}|\mathbf{x})$ and $p_{o}(\mathbf{u}|\mathbf{x})$  in the Hellinger metric. In this respect, the optimal data-driven DL detector derived from minimizing the KL divergence, $D_{\mathrm{KL}}(p_{o},p_{\boldsymbol{\theta}})$, also implies that $p_{\boldsymbol{\theta}}(\mathbf{u}|\mathbf{x})$ is in close proximity of $p_{o}(\mathbf{u}|\mathbf{x})$. 
	
	Let $\Theta_{R}=\{\boldsymbol{\theta}\,|\,\|\boldsymbol{\theta}\|_{\infty}\leq R, R\geq 1\}$ be the bounded subset of $\Theta$ and the performance of the data-driven DL detector will be evaluated within $\Theta_{R}$. Denote 
	\begin{align}
		J(p_{o})=\mathbb{E}\{\ln p_{o}(\mathbf{u}|\mathbf{x})\}
	\end{align}
	as the expectation of $\ln p_{o}(\mathbf{u}|\mathbf{x})$. The optimal data-driven DL detector that minimizes $D_{\mathrm{KL}}(p_{o},p_{\boldsymbol{\theta}})$ within $\Theta_{R}$ can be expressed as
	\begin{align}\label{mini1}
		\boldsymbol{\theta}_{o}&=\arg\min_{\boldsymbol{\theta}\in\Theta_{R}}D_{\mathrm{KL}}(p_{o},p_{\boldsymbol{\theta}})=\arg\max_{\boldsymbol{\theta}\in\Theta_{R}}J(p_{\boldsymbol{\theta}}),
	\end{align}
	where $J(p_{\boldsymbol{\theta}})=\mathbb{E}\{\ln p_{\boldsymbol{\theta}}(\mathbf{u}|\mathbf{x})\}$. Eq.\,\eqref{mini1} indicates that minimizing $D_{\mathrm{KL}}(p_{o},p_{\boldsymbol{\theta}})$ is equivalent to maximizing the log-likelihood $J(p_{\boldsymbol{\theta}})$.
	However, the optimization over $J(p_{\boldsymbol{\theta}})$ in~\eqref{mini1} is difficult to implement in practice. Generally,
	\begin{align}\label{empirical}
		J_{\mathcal{Z}}(p_{\boldsymbol{\theta}})=\frac{1}{|\mathcal{Z}|}\sum_{(\mathbf{x}_{m},\mathbf{u}_{m})\in\mathcal{Z}}\ln p_{\boldsymbol{\theta}}(\mathbf{u}_{m}|\mathbf{x}_{m})
	\end{align}
	is applied to optimize $\boldsymbol{\theta}$ with respect to (w.r.t.) $\mathcal{Z}$, and the corresponding maximum log-likelihood detector is given by
	\begin{align}\label{min2}
		\boldsymbol{\theta}_{\mathcal{Z}}=\arg\max_{\boldsymbol{\theta}\in\Theta_{R}}J_{\mathcal{Z}}(p_{\boldsymbol{\theta}}).
	\end{align}
	Obviously,
	\begin{align}
		\sum_{(\mathbf{x}_{m},\mathbf{u}_{m})\in\mathcal{Z}}\ln \frac{p_{\boldsymbol{\theta}_{\mathcal{Z}}}(\mathbf{u}_{m}|\mathbf{x}_{m})}{p_{\boldsymbol{\theta}}(\mathbf{u}_{m}|\mathbf{x}_{m})}\geq 0.
	\end{align}
	
	The detected symbol of the data-driven DL detector can then be expressed as
	\begin{align}\label{dl_det}
		\mathbf{s}_{\mathrm{DL}}=\arg\max_{\mathbf{s}\in\mathbb{S}^{d_{l+1}}} p_{\boldsymbol{\theta}_{\mathcal{Z}}}(\mathbf{f}_{\mathrm{map}}(\mathbf{s})|\mathbf{x}),
	\end{align} 
	where $\mathbf{f}_{\mathrm{map}}(\mathbf{s})$ is one-to-one mapping from $\mathbf{s}$ to $\mathbf{u}$. 
	
	According to~\eqref{kl-info}, $D_{\mathrm{KL}}(p_{o},p_{\boldsymbol{\theta}_{\mathcal{Z}}})$ can be decomposed into
	\begin{align}\label{decom}
		D_{\mathrm{KL}}(p_{o},p_{\boldsymbol{\theta}_{\mathcal{Z}}})=[J(p_{o})-J(p_{\boldsymbol{\theta}_{o}})]+[J(p_{\boldsymbol{\theta}_{o}})-J(p_{\boldsymbol{\theta}_{\mathcal{Z}}})].
	\end{align}
	
	The first term $[J(p_{o})-J(p_{\boldsymbol{\theta}_{o}})]$ in~\eqref{decom} is non-negative and is independent of $\mathcal{Z}$, referred to as the approximation error. The second term $[J(p_{\boldsymbol{\theta}_{o}})-J(p_{\boldsymbol{\theta}_{\mathcal{Z}}})]$ in~\eqref{decom} is also non-negative and is determined by $\mathcal{Z}$, called as the generalization error. 
	
	Denote $f(\mathbf{x})$ as an $\mathbb{R}^{d}\rightarrow\mathbb{R}$ function and $\ell_{2}$ be the finite 2-norm space of $f(\mathbf{x})$ with
	\begin{align}
		\|f(\mathbf{x})\|_{2}=\big[\mathbb{E}\{f^{2}(\mathbf{x})\}\big]^{1/2}<+\infty.
	\end{align}
	
	The following theorem, proved in Appendix~\ref{apex-a}, demonstrates that the approximation error in~\eqref{decom} can be narrowed down with any precision by ReLU DNNs.
	
	\begin{theorem}\label{theorem1}
		If $\ln p_{o}(\mathbf{u}|\mathbf{x})\in\ell_{2}$, then there exits an optimized DL estimator built on a ReLU DNN of $\boldsymbol{\theta}\in\Theta_{R}$ with sufficiently large $R$ and at most $	
		\lceil\log_{2}(d_{0}+1)\rceil
		$ hidden layers such that
		\begin{align}\label{theo1}
			J(p_{o})-J(p_{\boldsymbol{\theta}_{o}})\leq\varepsilon
		\end{align}
		for any $\varepsilon>0$.
	\end{theorem}
	
	\begin{remark}\label{remark3}
		From Theorem~\ref{theorem1}, the data-driven DL detector with bigger network size is more powerful at function representation and tends to have lower approximation error and better performance.
	\end{remark}
	\begin{remark}
		Theorem~\ref{theorem1} indicates that the data-driven DL detector is model independent and is capable of approximating any target \textit{posterior} distribution. Therefore, the data-driven DL detector is a preferred choice compared to other model-based MIMO detection algorithms if no specific channel model is known a priori or if complicated nonlinear systems are presented.
	\end{remark}

	Next, we will discuss the convergence of the generalization error in~\eqref{decom}. The following two auxiliary lemmas, proved in Appendixes~\ref{apex-b} and~\ref{apex-c}, respectively, are presented first.
	\begin{lemma}\label{lemma-1}
		Let $\alpha=R\|\mathbf{d}\|_{\infty}$, $\beta=\alpha/(\alpha-1)$,  and
		\begin{align}
			\nu=\mathbb{E}\big\{\big[(\ln d_{l+1}+1)(\alpha^{l+1}(\|\mathbf{x}\|_{2}+\beta)-\beta)\big]^{2}\big\}.
		\end{align} 
		Assume that $\mathbb{E}\{\|\mathbf{x}\|_{2}^{2}\}$ is finite and so is $\nu$. For any $\varepsilon>0$ and
		$|\mathcal{Z}|\geq 4\nu/\varepsilon^{2}$,
		we have
		\begin{align}\label{lemma1}
			\mathbf{P}\big(\sup_{\boldsymbol{\theta}\in\Theta_{R}}|J_{\mathcal{Z}}(p_{\boldsymbol{\theta}})-J(p_{\boldsymbol{\theta}})|>\varepsilon\big)\leq 4\mathbf{P}\big(\sup_{\boldsymbol{\theta}\in\Theta_{R}}|J_{\mathcal{Z}}^{\circ}(p_{\boldsymbol{\theta}})|>\frac{\varepsilon}{4}\big),
		\end{align}
		where $\mathbf{P}$ is the distribution of training samples in $\mathcal{Z}$ and 
		\begin{align}
			J_{\mathcal{Z}}^{\circ}(p_{\boldsymbol{\theta}})=\frac{1}{|\mathcal{Z}|}\sum_{m=1}^{|\mathcal{Z}|}\omega_{m}\ln p_{\boldsymbol{\theta}}(\mathbf{u}_{m}|\mathbf{x}_{m})
		\end{align} 
		with $\{\omega_{1},\ldots,\omega_{|\mathcal{Z}|}\}$ a Rademacher sequence.
	\end{lemma}
	
	\begin{lemma}\label{lemma2}
		Assume that $\frac{1}{|\mathcal{Z}|}\sum_{m=1}^{|\mathcal{Z}|}\|\mathbf{x}_{m}\|_{2}^{2}\leq\delta^{2}$. For any $\boldsymbol{\theta},\boldsymbol{\lambda}\in\Theta_{R}$, we have
		\begin{align}
			&|J_{\mathcal{Z}}(p_{\boldsymbol{\theta}})-J_{\mathcal{Z}}(p_{\boldsymbol{\lambda}})|\leq 3^{l+1}\|\mathbf{d}\|_{\infty}\alpha^{l}(\delta+\beta)\|\boldsymbol{\theta}-\boldsymbol{\lambda}\|_{\infty}.
		\end{align}	
	\end{lemma}

	If there exits a collection of functions $\mathbf{p}_{1}(\mathbf{x}),\ldots,\mathbf{p}_{C}(\mathbf{x})\in\mathcal{D}$ with their parameters belonging to $\Theta_{R}$ for $C\in\mathbb{N}$, the functions in this collection satisfy
	\begin{align}\label{cover}
		|J_{\mathcal{Z}}(p_{\boldsymbol{\theta}})-J_{\mathcal{Z}}(p_{j})|\leq\varepsilon,\ \forall j\in\{1,\ldots,C\},
	\end{align}
	for any $\mathbf{p}_{\boldsymbol{\theta}}(\mathbf{x})\in\mathcal{D}$ with $\boldsymbol{\theta}\in\Theta_{R}$ and $\varepsilon>0$, where
	\begin{align}
		p_{j}(\mathbf{u}|\mathbf{x})=\prod_{i=1}^{d_{l+1}}p_{j,i}(\mathbf{x})^{\mathbf{1}_{\tilde{\mathbf{u}}_{i}}(\mathbf{u})}
	\end{align}
	and $p_{j,i}(\mathbf{x})$ is the $i$-th entry of $\mathbf{p}_{j}(\mathbf{x})$.
	
	Define the covering number $C(\varepsilon, \Theta_{R})$ as the smallest value of $C\in\mathbb{N}$ that satisfies~\eqref{cover}. According to Lemma~\ref{lemma2} and~\cite[Lemma~3]{qiang2020channel}, the upper bound on $\ln C(\varepsilon, \Theta_{R})$ is given by
	\begin{align}\label{covering}
		\ln C(\varepsilon, \Theta_{R})\leq d_{s}\ln\Big[\frac{3^{l+1}4\|\mathbf{d}\|_{\infty}\alpha^{l}(\delta+\beta)}{\varepsilon}\Big]
	\end{align}
	for any $\varepsilon>0$.

	Following~\eqref{covering}, the next theorem, proved in Appendix~\ref{apex-d}, demonstrates the rate of convergence of the generalization error in~\eqref{decom}.
	
	\begin{theorem}\label{theorem2}
		Let $\alpha=R\|\mathbf{d}\|_{\infty}$, $\beta=\alpha/(\alpha-1)$, $\mu=\mathbb{E}\{\|\mathbf{x}\|_{2}^{2}\}$, and
		\begin{align}
			\nu=\mathbb{E}\big\{\big[(\ln d_{l+1}+1)(\alpha^{l+1}(\|\mathbf{x}\|_{2}+\beta)-\beta)\big]^{2}\big\}.
		\end{align} Denote $\sigma$ as the variance of $\|\mathbf{x}\|_{2}^{2}$. Let $\delta_{1}=[(\ln d_{l+1}+1)(\alpha^{l+1}\delta+\beta)-\beta)]^{2}$ and $\delta_{2}=3^{l+1}2^{6}\alpha^{l}(\delta+\beta)$ for any $\delta^{2}\geq\mu$. For any $\varepsilon>0$, we have
		\begin{align}
			\mathbf{P}([J(p_{\boldsymbol{\theta}_{o}})-J(p_{\boldsymbol{\theta}_{\mathcal{Z}}})]>\varepsilon)\leq 8\mathrm{exp}\Big(-\frac{|\mathcal{Z}|\varepsilon^{2}}{1024\delta_{1}}\Big)+\frac{4\sigma^{2}}{|\mathcal{Z}|(\delta^{2}-\mu)^{2}}
		\end{align}
		if $|\mathcal{Z}|\geq 16\nu/\varepsilon^{2}$ and $|\mathcal{Z}|\geq (1024\delta_{1}d_{s}\ln\frac{\delta_{2}}{\varepsilon})/\varepsilon^{2}$.
	\end{theorem}
	\begin{remark}
		Though the approximation error can be reduced by enlarging network size as indicated in Theorem~\ref{theorem1}, Theorem~\ref{theorem2} demonstrates that the rate of the convergence of the generalization error will decrease. Therefore, a tradeoff between the generalization error and the approximation error should be carefully balanced when implementing the data-driven DL detector.	
	\end{remark}
	\begin{remark}
		Theorem~\ref{theorem2} demonstrates that the rate of convergence of $J(p_{\boldsymbol{\theta}_{\mathcal{Z}}})$ to $J(p_{\boldsymbol{\theta}_{o}})$ grows at least polynomially with $|\mathcal{Z}|$ under a fixed network structure.
		
	\end{remark}

	The following corollary, proved in Appendix~\ref{apex-e}, presents our main conclusion on the performance of the data-driven DL detector.
	\begin{corollary}\label{corollary}
		For any $\varepsilon>0$ and sufficiently large $R$, there exits a DL detector built on a ReLU DNN with at most
		$\lceil\log_{2}(d_{0}+1)\rceil
		$ hidden layers and $\boldsymbol{\theta}\in\Theta_{R}$
		such that
		\begin{align}\label{coro1-1}
			\lim_{|\mathcal{Z}|\rightarrow+\infty}\mathbf{P}(D_{\mathrm{KL}}(p_{o},p_{\boldsymbol{\theta}_{\mathcal{Z}}})>\varepsilon)=0.
		\end{align}
	\end{corollary}

	\begin{remark}\label{remark2}
		Corollary~\ref{corollary} and~\eqref{helling} show that
		\begin{align}\label{con}
			p_{\boldsymbol{\theta}_{\mathcal{Z}}}(\mathbf{u}|\mathbf{x})\approx p_{o}(\mathbf{u}|\mathbf{x}),
		\end{align}
		if $|\mathcal{Z}|$ is sufficiently large and the network structure is suitably configured. Therefore, the data-driven DL detector can learn to fit $p_{o}(\mathbf{u}|\mathbf{x})$ perfectly and is very suitable for communication systems with unknown nonlinear detrimental effects.			
	\end{remark}
	
	\subsection{Performance Analysis with CSI}\label{incomplete}
	
	\begin{figure*}[t]
		\vskip 0.2in
		\centering
		%\hspace{3.1em}
		\subfigure[]{
			\begin{minipage}[t]{0.46\linewidth}
				\centering
				\includegraphics[width=\columnwidth]{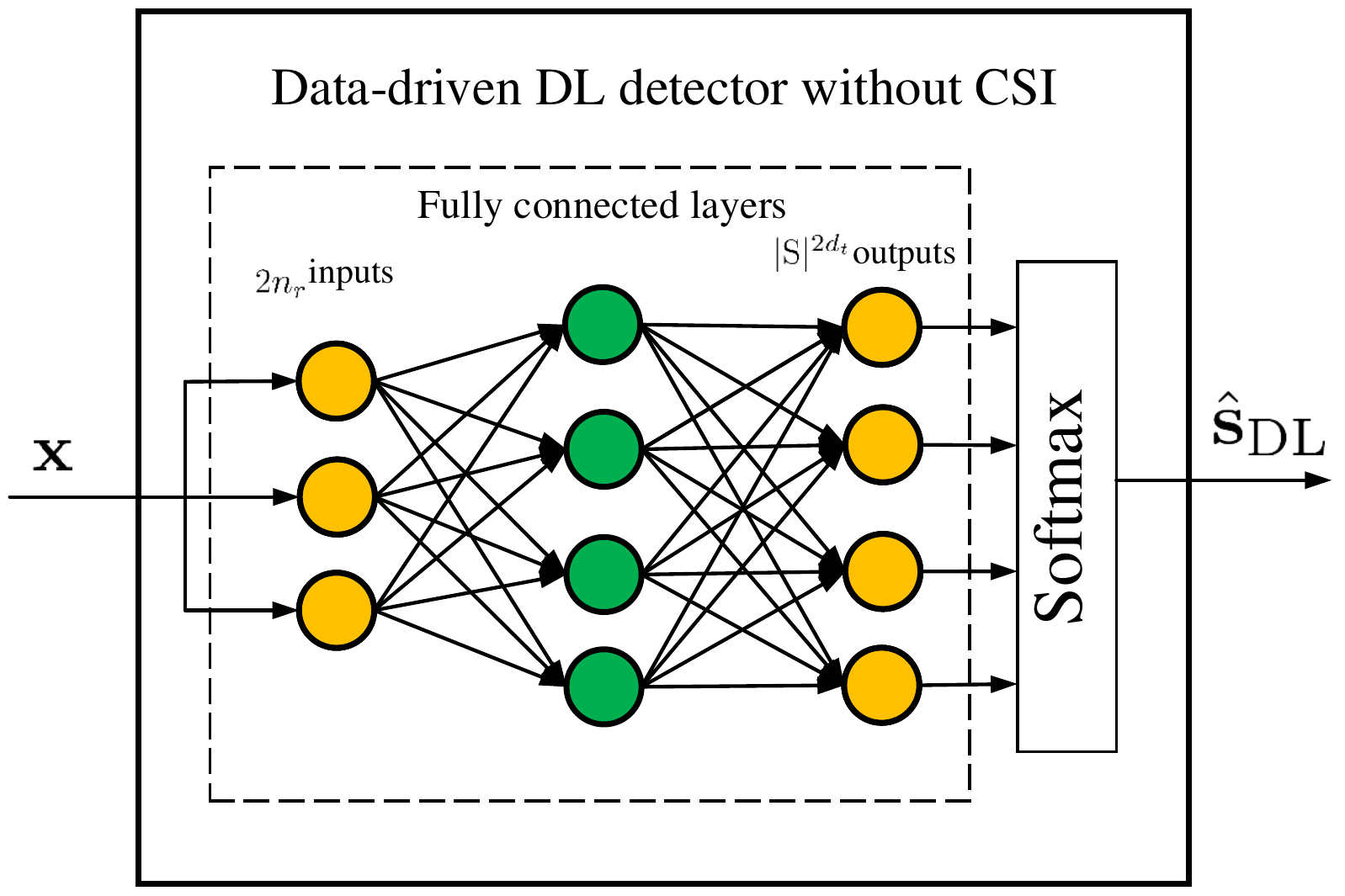}
				%\caption{Training data of Circle}
			\end{minipage}
		}
		%\hspace{1.4em}
		\subfigure[]{
			\begin{minipage}[t]{0.46\linewidth}
				\centering
				\includegraphics[width=\columnwidth]{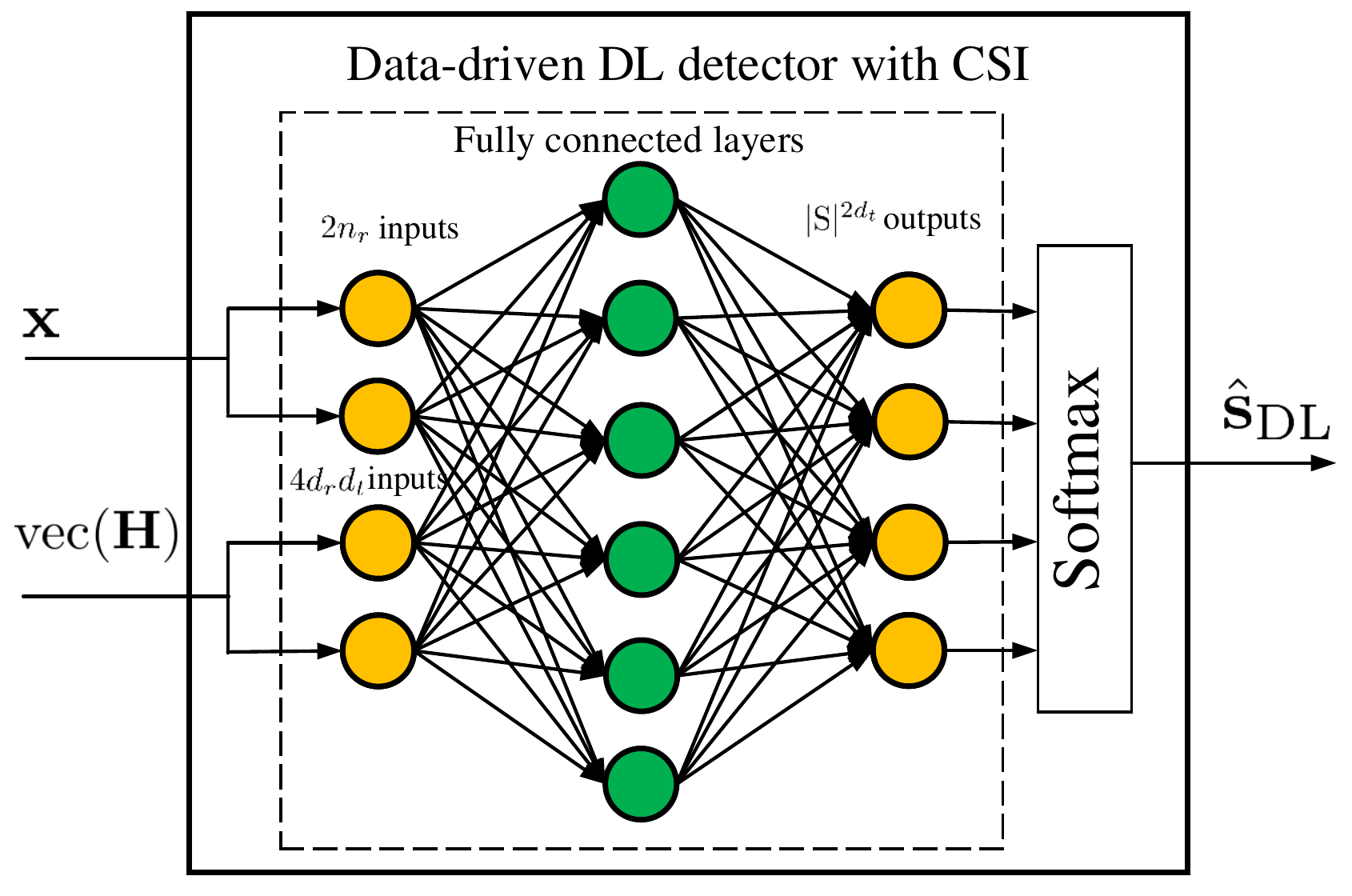}
				%\caption*{(b)}
			\end{minipage}
		}

		\caption{Network structures for the data-driven DL detectors with CSI and without CSI.}
		\label{figure6-2}
		\vskip -0.2in
	\end{figure*}

	The data-driven DL detector derived in~\eqref{dl_det} is model independent and requires no CSI to learn a detection mapping approaching MAP performance. However, it is only applicable when the channel is static and it does not perform well for varying channels~\cite{Samuel2019Learning}. Next, we will show the necessity of CSI for DL-based MIMO detection to generalize over the whole distribution of $\mathbf{H}$.
	
	From~\eqref{con}, the data-driven detector takes the \textit{posterior} probability, $p_{o}(\mathbf{u}|\mathbf{x})$, as the target and we have
	\begin{align}\label{ml1}
		p_{o}(\mathbf{u}|\mathbf{x})=p_{o}(\mathbf{u}|\mathbf{f}_{\mathrm{nlr}}(\mathbf{H}\mathbf{s}+\mathbf{n}))
	\end{align}
	and 
	\begin{align}
		p_{o}(\mathbf{u}|\mathbf{x})=p_{o}(\mathbf{u}|(\mathbf{H}\mathbf{f}_{\mathrm{nlt}}(\mathbf{s})+\mathbf{n}))
	\end{align}
	according to~\eqref{nonlinear1} and~\eqref{nonlinear2}, respectively. To exploit $\mathbf{s}_{\mathrm{DL}}$ in different scenarios, we discuss two channel states:
	\begin{itemize}
		\item In time-invariant channels, $\mathbf{H}$ remains deterministic and constant during the training and detecting phases.
		\item In time-varying channels, $\mathbf{H}$ is randomly generated from a known continuous distribution and changes in each realization of one training sample.
	\end{itemize}
	
	In the time-invariant channels, channel matrices in training data and deployed environment are identical, and $\mathbf{s}$ can be recovered with low detection error using the minimum distance rule. In view of  $\mathbf{s}_{\mathrm{DL}}\approx\mathbf{s}_{\mathrm{MAP}}$ from Corollary~\ref{corollary}, the performance of the data-driven DL detector without CSI is not affected by the state of $\mathbf{H}$ in the time-invariant channels.

	In time-varying channels, channel matrices in training data and deployed environment are different due to randomness. As a consequence, $p_{\boldsymbol{\theta}_{\mathcal{Z}}}(\mathbf{u}|\mathbf{x})$ learned from training data is inconsistent with the real \textit{posterior} probability in the deployed environment and $\mathbf{s}$ is indistinguishable using the maximum likelihood (ML) rule. The data-driven DL detector is unable to decouple transmitted symbols from time-varying channels owing to a lack of CSI and using a single network without the information of $\mathbf{H}$ cannot generalize the entire distribution of possible channels for MIMO detection.
	
	To alleviate the impact of time-varying channels, let us consider the case that channel $\mathbf{H}$ is already known. The MAP detector in~\eqref{map_detec} is converted into
	\begin{align}\label{map1}
		\mathbf{s}_{\mathrm{MAP}}=\arg\max_{\mathbf{s}\in\mathbb{S}^{2d_{t}}}p_{o}(\mathbf{s}|\,\mathbf{H},\mathbf{x}),
	\end{align}
	where $p_{o}(\mathbf{s}|\,\mathbf{H},\mathbf{x})$ is the \textit{posterior} probability of $\mathbf{s}$ given $\mathbf{H}$ and $\mathbf{x}$. 
	
	To approximate the MAP detector in~\eqref{map1}, the data-driven DL detector should take both $\mathbf{H}$ and $\mathbf{x}$ as its input and other structures of the data-driven DL detector remain the same. The network structures for the data-driven DL detectors with and without CSI are illustrated in Fig.~\ref{figure6-2}. Similar to Corollary~\ref{corollary}, we can also prove that the data-driven DL detector with CSI can well approximate the MAP detector in~\eqref{map1}. Therefore, the data-driven DL detector manages to generalize over all possible realizations of $\mathbf{H}$ by incorporating CSI into the network design.
	\begin{remark}
		
		In time-varying channels, CSI is essential to the data-driven DL detector to detect $\mathbf{s}$ over all possible realizations of the channel. Moreover, the data-driven DL detector with CSI is also model independent and can achieve the MAP comparable performance over various scenarios. In this situation, the data-driven DL detector tends to have a large network structure by taking $\mathbf{H}$ as the input and thus requires enormous train samples to converge, as shown in Theorem~\ref{theorem2}. The data-driven DL detector with CSI yields optimal accuracy but is prohibitive for large-sized MIMO systems.
	\end{remark}

	\section{Model-driven DL Detector}\label{model-driven} 
	
	Simply implementing the DL detector in a data-driven fashion is ineffective in some practical scenarios, especially for large-sized MIMO systems. An alternative way is to integrate the model expert knowledge into the network structure and design model-driven DL detectors. Instead of using a conventional DNN, model-driven DL detectors usually associate iterative detection algorithms with DL, which can mitigate the impact of time-varying channels, accelerate the convergence, and reduce the complexity~\cite{hedetec2020}. Each layer of the model-driven DL detector operates a single iteration in~\eqref{iterative} and adds some trainable variables to enhance the detection performance. 
	
	Suppose that the MIMO channel model is linear as shown in~\eqref{signal_model2} and channel $\mathbf{H}$ is known at the receiver. Model-driven DL detector $\mathcal{D}_{u}$ is based on the iteration detector in~\eqref{iterative}  and is given by
	\begin{align}
		\mathbf{f}_{\boldsymbol{\vartheta}}(\mathbf{H},\mathbf{x})=\tilde{\mathcal{A}}_{\boldsymbol{\vartheta}_{l_{u}}}\circ\tilde{\mathcal{A}}_{\boldsymbol{\vartheta}_{l_{u}-1}}\circ\cdots\circ\tilde{\mathcal{A}}_{\boldsymbol{\vartheta}_{0}},
	\end{align} 
	where $\boldsymbol{\vartheta}_{i}$ is the trainable parameter, $\boldsymbol{\vartheta}=(\boldsymbol{\vartheta}_{0},\ldots,\boldsymbol{\vartheta}_{l_{u}})$ is the parameter of $\mathcal{D}_{u}$, and $\tilde{\mathcal{A}}_{\boldsymbol{\vartheta}_{i}}$ is the computation at the $i$-th iteration that is parameterized by $\boldsymbol{\vartheta}_{i}$ for $i\in\{0,\ldots,l_{u}\}$. In particular, $\tilde{\mathcal{A}}_{\boldsymbol{\vartheta}_{i}}$ depends on underlying iterative detection algorithm, i.e., $\mathbf{f}_{\mathrm{iter}}$ in~\eqref{iterative}. The training sample set of $\mathcal{D}_{u}$ is defined by
	\begin{align}
		\Omega=\{(\mathbf{x}_{m},\mathbf{H}_{m},\mathbf{s}_{m})|\mathbf{x}\in\mathbb{R}^{2d_{r}},\mathbf{H}\in\mathbb{R}^{2d_{r}\times 2d_{t}},\mathbf{s}\in\mathbb{S}^{2d_{t}},m=1,\ldots,|\Omega|\}.
	\end{align}
	
	Let $J(\mathbf{f}_{\boldsymbol{\vartheta}}(\mathbf{x},\mathbf{H}))$ be the non-negative loss function between $\mathbf{f}_{\boldsymbol{\vartheta}}(\mathbf{x},\mathbf{H})$ and $\mathbf{s}$ and $\Theta_{u}=\{\boldsymbol{\vartheta}\in\mathbb{R}^{\tilde{d}_{u}}\}$ be the parameter set for $\tilde{d}_{u}\in\mathbb{N}$.\footnote{The selection of $J(\mathbf{f}_{\boldsymbol{\vartheta}}(\mathbf{x},\mathbf{H}))$ is not fixed, which may be MSE or KL divergence.} The goal of $\mathcal{D}_{u}$ is to optimize $\boldsymbol{\vartheta}$ by minimizing $J(\mathbf{f}_{\boldsymbol{\vartheta}})$ within $\Theta_{u}$,
	\begin{align}
		\boldsymbol{\vartheta}_{o}=\arg\min_{\boldsymbol{\vartheta}\in\Theta_{u}}\mathbb{E}\{J(\mathbf{f}_{\boldsymbol{\vartheta}}(\mathbf{x},\mathbf{H}))\}=\arg\min_{\boldsymbol{\vartheta}\in\Theta_{u}}J_{u}(\mathbf{f}_{\boldsymbol{\vartheta}}),
	\end{align}
	where $\boldsymbol{\vartheta}_{o}$ is the optimized parameter and $J_{u}(\mathbf{f}_{\boldsymbol{\vartheta}_{o}})$ has the lowest mean loss over all $\boldsymbol{\vartheta}\in\Theta_{u}$.
	
	However, it is difficult to obtain the explicit form of $J_{u}(\mathbf{f}_{\boldsymbol{\vartheta}})$ and the empirical mean
	\begin{align}
		J_{\Omega}(\mathbf{f}_{\boldsymbol{\vartheta}})=\frac{1}{|\Omega|}\sum_{(\mathbf{x}_{m},\mathbf{H}_{m},\mathbf{h}_{m},)\in\Omega}J(\mathbf{f}_{\boldsymbol{\vartheta}}(\mathbf{x}_{m},\mathbf{H}_{m}))
	\end{align}
	is typically used to optimize $\boldsymbol{\vartheta}$ w.r.t. $\Omega$. Denote 
	\begin{align}\label{unfolded}
		\boldsymbol{\vartheta}_{\Omega} = \arg\min_{\boldsymbol{\vartheta}\in\Theta_{u}}J_{\Omega}(\mathbf{f}_{\boldsymbol{\vartheta}}),
	\end{align}
	and
	\begin{align}\label{unfolded3}
		J_{\Omega}(\mathbf{f}_{\boldsymbol{\vartheta}_{\Omega}})=\min_{\boldsymbol{\vartheta}\in\Theta_{u}}J_{\Omega}(\mathbf{f}_{\boldsymbol{\vartheta}_{\Omega}}).
	\end{align}
	From~\eqref{unfolded} and~\eqref{unfolded3}, $\mathbf{f}_{\boldsymbol{\vartheta}_{\Omega}}(\mathbf{x},\mathbf{H})$ is the optimal model-driven DL detector trained by $\Omega$. Therefore, we can evaluate the performance of the model-driven DL detector by quantifying the distance between $J_{u}(\mathbf{f}_{\boldsymbol{\vartheta}_{\Omega}})$ and $J_{u}(\mathbf{f}_{\boldsymbol{\vartheta}_{o}})$.
	
	For an integer $C\in\mathbb{N}$ and a collection of functions $\mathbf{f}_{1}(\mathbf{x},\mathbf{H}),\ldots,\mathbf{f}_{C}(\mathbf{x},\mathbf{H})\in\mathcal{D}_{u}$, let $C_{u}(\epsilon,\Theta_{u})$ be the smallest value of $C$ such that
	\begin{align}
		\min_{j\in\{1,\ldots,C\}}\big|J_{\Omega}(\mathbf{f}_{\boldsymbol{\vartheta}})-J_{\Omega}(\mathbf{f}_{j})\big|\leq\varepsilon
	\end{align}
	for $\varepsilon>0$ and any $\mathbf{f}_{\boldsymbol{\vartheta}}(\mathbf{x},\mathbf{H})\in\mathcal{D}_{u}$. The following theorem, proved in Appendix~\ref{apex-f}, shows the relationship between $J_{u}(\mathbf{f}_{\boldsymbol{\vartheta}_\Omega})$ and $J_{u}(\mathbf{f}_{\boldsymbol{\vartheta}_o})$.
	\begin{theorem}\label{theorem3}
		For any $\varepsilon>0$ and $\delta_{u}>0$, we have
		\begin{align}\label{unfolded_network}
			\mathbf{P}_{u}([J_{u}(\mathbf{f}_{\boldsymbol{\vartheta}_\Omega})-J_{u}(\mathbf{f}_{\boldsymbol{\vartheta}_o})]>\varepsilon
			)\leq8\exp\left(\ln C_{u}-\frac{|\Omega|\varepsilon^{2}}{512\delta_{u}}\right)+\mathbf{P}_{\Omega},
		\end{align}
		where $\mathbf{P}_{u}$ denotes the distribution of the training samples in $\Omega$ and 
		\begin{align}
			\mathbf{P}_{\Omega}=\mathbf{P}_{u}\bigg(\frac{1}{|\Omega|}\sum_{m=1}^{|\Omega|}[J(\mathbf{f}_{\boldsymbol{\vartheta}}(\mathbf{x}_{m},\mathbf{H}_{m}))]^{2}\geq\delta_{u}\bigg).
		\end{align}
	\end{theorem}
	
	In general, $\frac{1}{|\Omega|}\sum_{m=1}^{|\Omega|}[J(\mathbf{f}_{\boldsymbol{\vartheta}}(\mathbf{x}_{m},\mathbf{H}_{m}))]^{2}$ converges to its mean value as $|\Omega|$ increases, and $|J_{u}(\mathbf{f}_{\boldsymbol{\vartheta}_\Omega})-J_{u}(\mathbf{f}_{\boldsymbol{\vartheta}_o})|$ also converges to zero if $\ln C_{u}$ in~\eqref{unfolded_network} is finite. Therefore, $\mathbf{f}_{\boldsymbol{\vartheta}_{\Omega}}(\mathbf{x},\mathbf{H})$, asymptotically approaches to the optimal model-driven DL detector,  $\mathbf{f}_{\boldsymbol{\vartheta}_{o}}(\mathbf{x},\mathbf{H})$, as the size of training data increases. Furthermore, the model-driven DL detector $\mathbf{f}_{\boldsymbol{\vartheta}}(\mathbf{x},\mathbf{H})$ is equivalent to underlying iterative detection algorithm $\mathbf{f}(\mathbf{x},\mathbf{H})$ if $\boldsymbol{\vartheta}$ is set to be specific value so that $\hat{\mathcal{A}}_{\boldsymbol{\vartheta}_{i}}=\hat{\mathcal{A}}_{i}$ for $i\in\{0,\ldots,l_{u}\}$. Hence, the performance of the model-driven DL detectors is better or at least equal to that of its underlying iterative detection algorithm, as indicated by Theorem~\ref{theorem3}.
	\begin{remark}
		In general, the dimension of the parameter space of the model-driven detector, $\tilde{d}_{u}$, is much smaller than that of the data-driven DL detector. Theorem~\ref{theorem3} demonstrates that model-driven DL detectors require far fewer train samples to converge, making them more suitable to large-sized MIMO systems. The performance of the model-driven DL detectors is determined by underlying iterative detection algorithms, while most of these algorithms cannot guarantee the convergence to the MAP detector except under some special scenarios. Generally, there exits the performance gap between the model-driven and the data-driven DL detectors.	
	\end{remark}

	\begin{remark} 
		The model-driven DL detectors are customized for specific MIMO systems and thus do not capture the model independent property of DL. On the other hand, the model-driven DL detectors may become unreliable and divergent where the presumptive system model mismatches the real environment, which severely limits their applications.
	\end{remark}

	\section{Simulation Results}\label{simulation}
	
	In this section, computer simulation is provided to verify that the data-driven DL detector asymptotically approaches to the MAP detector under linear and nonlinear MIMO systems. Moreover, simulation results shows that CSI is essential for the data-driven DL detector to achieve the MAP comparable performance over the time-varying channels. Moreover, simulation results demonstrate that the underlying iterative detection algorithm is the determinant factor that affects the performance of the model-driven DL detector.

	\subsection{Simulation Setting}
	The SNR is defined as
	\begin{align}
		\mathrm{SNR} = \frac{\mathbb{E}\|\mathbf{H}\mathbf{s}\|_{2}^{2}}{\mathbb{E}\|\mathbf{n}\|_{2}^{2}}.
	\end{align}
	The network of the data-driven DL detector has $4$ layers and each hidden layer is equipped with the same number of neurons. Denote $\tilde{d}$ as the width of the data-driven DL detector and $\tilde{d}=100$. We consider QPSK constellation and all transmitted symbols are generated with equal probability. 
	
	For data-driven DL detector, the simulation results are evaluated under the FC and VC cases, respectively. Unless stated otherwise, we train the network of the data-driven DL detector for $1,00,000$ iterations with a batch size of $2,000$ independently generated samples and test over $10,000$ samples. The MAP detector in~\eqref{map_detec} serves as the benchmark. The traditional model-based ZF, AMP, and SD algorithms are used to test against the data-driven DL detector.
	
	\begin{figure}[!t]
		\centering
		\includegraphics[width=0.6\columnwidth]{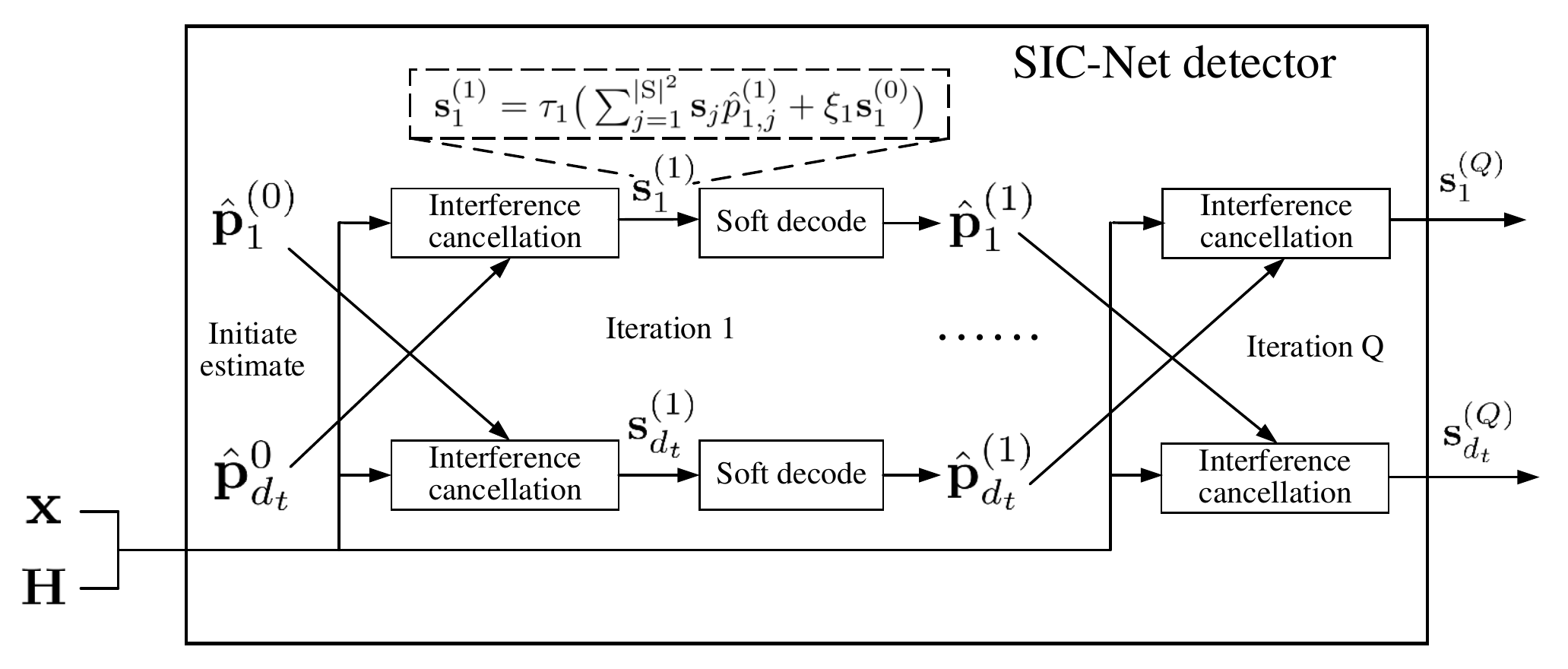}
		%where an .eps filename suffix will be assumed under latex, 
		% and a .pdf suffix will be assumed for pdflatex; or what has been declared
		% via \DeclareGraphicsExtensions.
		\caption{The network structure of the SIC-Net.}
		\label{figure6-3}
	\end{figure}
	
	The model-driven DL detector in simulation is based on the SIC algorithm, which is referred as to SIC-Net. Let $\hat{\mathbf{p}}_{i}^{(q)}\in\mathbb{R}^{|\mathbb{S}|^{2}}$ denote the estimated probability vector at the $i$-th transmitted antenna for $i\in\{1,\ldots,d_{t}\}$, where $q\in\{1,\ldots,Q\}$ is the iteration index for $Q\in\mathbb{N}$. The $j$-th entry of $\hat{\mathbf{p}}_{i}^{(q)}\in\mathbb{R}^{|\mathbb{S}|^{2}}$ is $\hat{p}_{i,j}^{(q)}$ that presents the estimated probability of $\mathbf{s}_{j}$ for $j\in\{1,\ldots,|\mathbb{S}|^{2}\}$. Let $\mathbf{s}_{i}^{(q)}$ be the estimated expected symbol at the $i$-th transmitted antenna for the $q$-th iteration, computed via
	\begin{align}\label{unfolded2}
		\mathbf{s}_{i}^{(q)}=\tau_{q}\bigg(\sum_{j=1}^{|\mathbb{S}|^{2}}\mathbf{s}_{j}\hat{p}_{i,j}^{(q)}+\xi_{q}\mathbf{s}_{i}^{(q-1)}\bigg),
	\end{align}
	where $\tau_{q}$ and $\xi_{q}$ are trainable variables. Other settings are the same as the SIC algorithm in~\cite{choi2000} and the corresponding network structure is illustrated in Fig.~\ref{figure6-3}. The performance of SIC-Net is evaluated in linear Gaussian MIMO channels. We train SIC-Net with a relatively small $5,000$ samples and test over $10,000$ samples. The MAP detector in~\eqref{map_detec} is used as the benchmark. Other model-driven DetNet~\cite{Samuel2019Learning} and OAMP-Net~\cite{hedetec2020} MIMO detectors are adopted for comparison.

	\subsection{Linear Systems}\label{linear-simu}
	In this subsection, we investigate the bit-error rate (BER) performance and convergence of the data-driven DL detector under a linear MIMO model in~\eqref{signal_model2}. 
	\begin{figure}[!t]
		\centering
		\includegraphics[width=0.6\columnwidth]{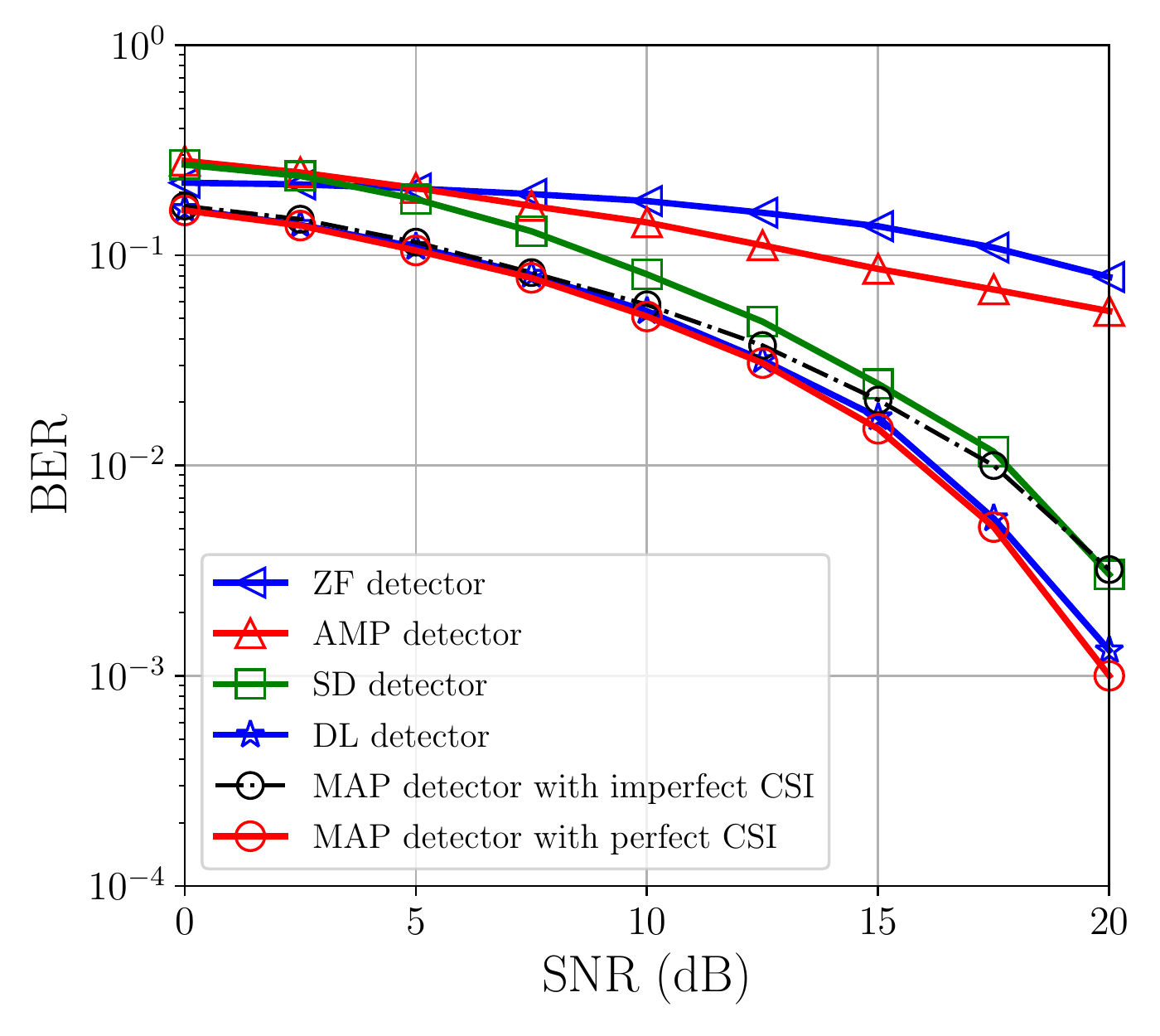}
		%where an .eps filename suffix will be assumed under latex, 
		% and a .pdf suffix will be assumed for pdflatex; or what has been declared
		% via \DeclareGraphicsExtensions.
		\caption{The BER performance of the data-driven DL detector versus SNR compared to other MIMO detectors over the time-invariant channel.}
		\label{figure3-1}
	\end{figure}
	
	\begin{figure*}[t]
		\vskip 0.2in
		\centering
		%\hspace{3.1em}
		\subfigure[Network width]{
			\begin{minipage}[t]{0.46\linewidth}
				\centering
				\includegraphics[width=\columnwidth]{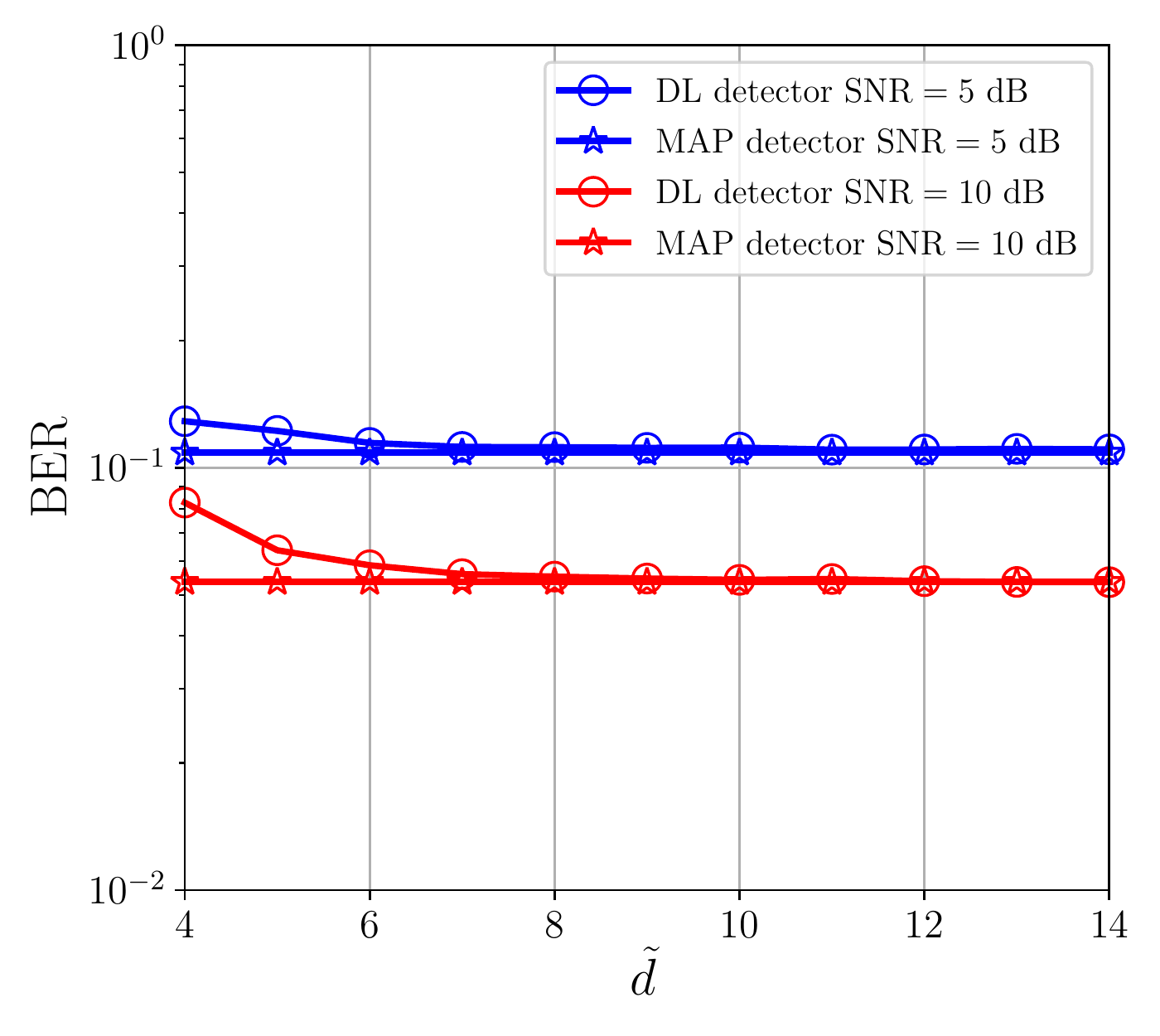}
				%\caption{Training data of Circle}
			\end{minipage}
		}
		%\hspace{1.4em}
		\subfigure[Sample size]{
			\begin{minipage}[t]{0.46\linewidth}
				\centering
				\includegraphics[width=\columnwidth]{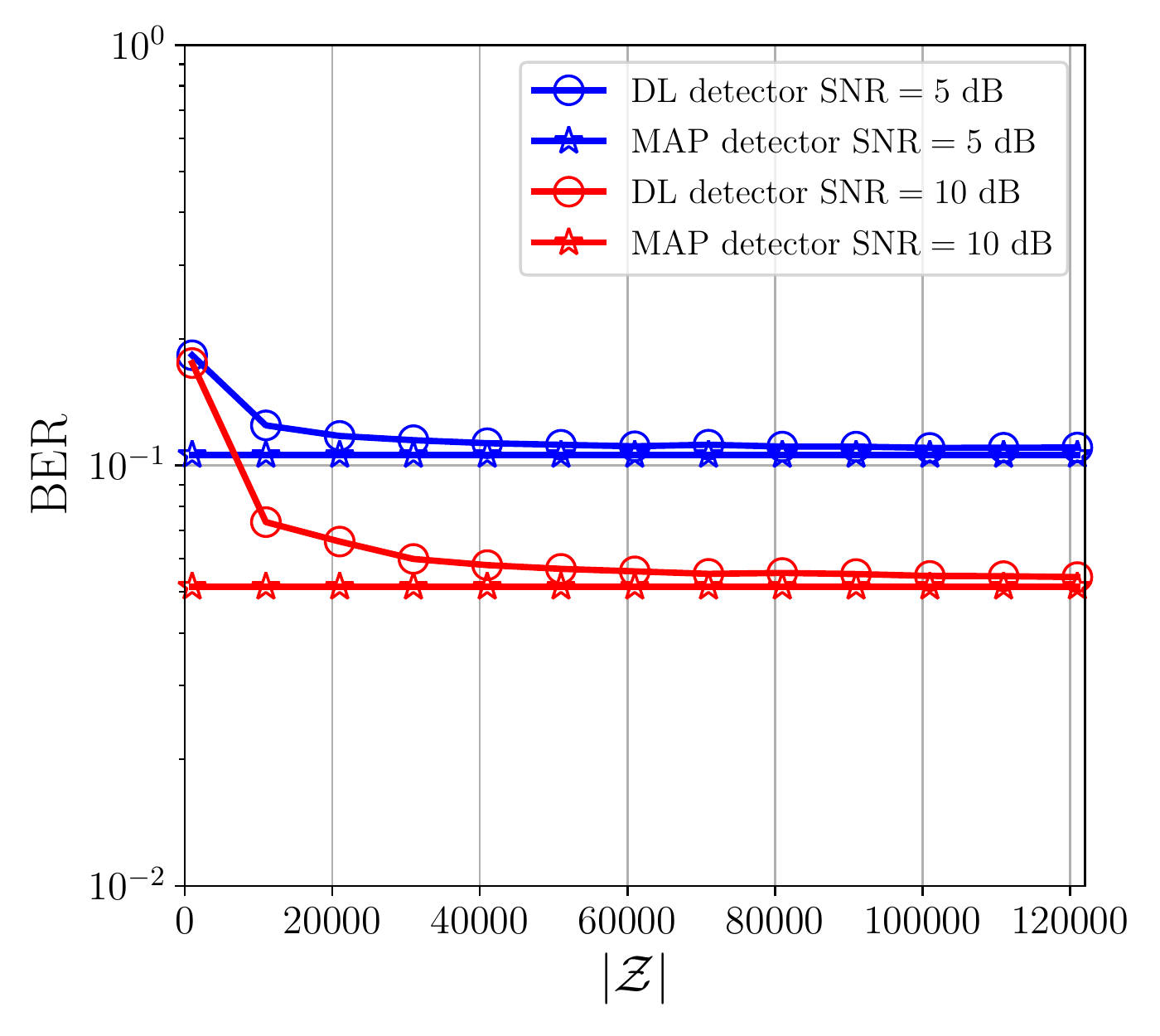}
				%\caption*{(b)}
			\end{minipage}
		}

		\caption{The BER performance of the data-driven DL detector versus $\tilde{d}$ and $|\mathcal{Z}|$ over the time-invariant channel, respectively.}
		\label{figure2}
		\vskip -0.2in
	\end{figure*}

	\subsubsection{Time-invariant Channel}\label{FC Model}
	
	Fig.~\ref{figure3-1}(a) compares the BER performance of the model-based ZF, AMP, SD, MAP and the data-driven DL detectors versus SNR where a $4\times 4$ time-invariant correlated channel is generated according to the one-ring model in~\cite{shiu2000}. We assume that perfect CSI is available for the ZF, AMP, and SD detectors while data-driven DL detector has no CSI. Moreover, the MAP detector is evaluated under both perfect and imperfect CSI, respectively. As shown in Fig.~\ref{figure3-1}, the data-driven DL detector can well approximate the MAP detector of perfect CSI and significantly outperforms other model-based detectors, which confirms that $p_{\boldsymbol{\theta}_{\mathcal{Z}}}(\mathbf{u}|\mathbf{x})\approx p_{o}(\mathbf{u}|\mathbf{x})$ in Corollary~\ref{corollary}. Moreover, the MAP detector with imperfect CSI suffers from serious performance degradation. Nevertheless, the data-driven DL detector is immune to CSI uncertainty since it requires no channel information for training. 
	
	Fig.~\ref{figure2}(a) shows the BER performance of the data-driven DL detector versus the network width, $\tilde{d}$, under fixed SNRs with the $2\times 2$ Gaussian channel. We train the network for $400,000$ independently generated samples. The BERs of the MAP detectors derived at the same SNRs are used as the benchmark. The approximation error determines the BER performance of the DL estimator since $|\mathcal{Z}|$ is sufficiently large. When $\tilde{d}$ is small, the dimension of the parameter space $\Theta_{R}$ is not big enough to fit with $p_{o}(\mathbf{u}|\mathbf{x})$. Hence, the BERs of the data-driven DL detector is significantly larger than those of the MAP detector. As $\tilde{d}$ increases, the dimension of the parameter space of $\Theta_{R}$ is enlarged and the approximation error decreases until both BERs converge, which verifies Theorem~\ref{theorem1}.

	Fig.~\ref{figure2}(b) shows the BER performance of the data-driven DL estimator versus the size of training samples, $|\mathcal{Z}|$, over the $2\times 2$ Gaussian channel. The SNRs are fixed and $\tilde{d}=100$. As in Fig.~\ref{figure2}(a), the BERs of the MAP detector are used as the benchmark. Similarly, the generalization error is the main factor that affects the BER performance of the DL estimator under large $\tilde{d}$. When $|\mathcal{Z}|$ is small, the BERs of the data-driven DL detector do not converge and are significantly larger than those of the MAP estimator. As $|\mathcal{Z}|$ increases, the BERs of the data-driven DL detector gradually approach to those of the MAP detector, which verifies Theorem~\ref{theorem2}.

	\subsubsection{Time-varying Channel} 
\begin{figure*}[t]
		\vskip 0.2in
		\centering
		%\hspace{3.1em}
		\subfigure[Gaussian channel]{
			\begin{minipage}[t]{0.46\linewidth}
				\centering
				\includegraphics[width=\columnwidth]{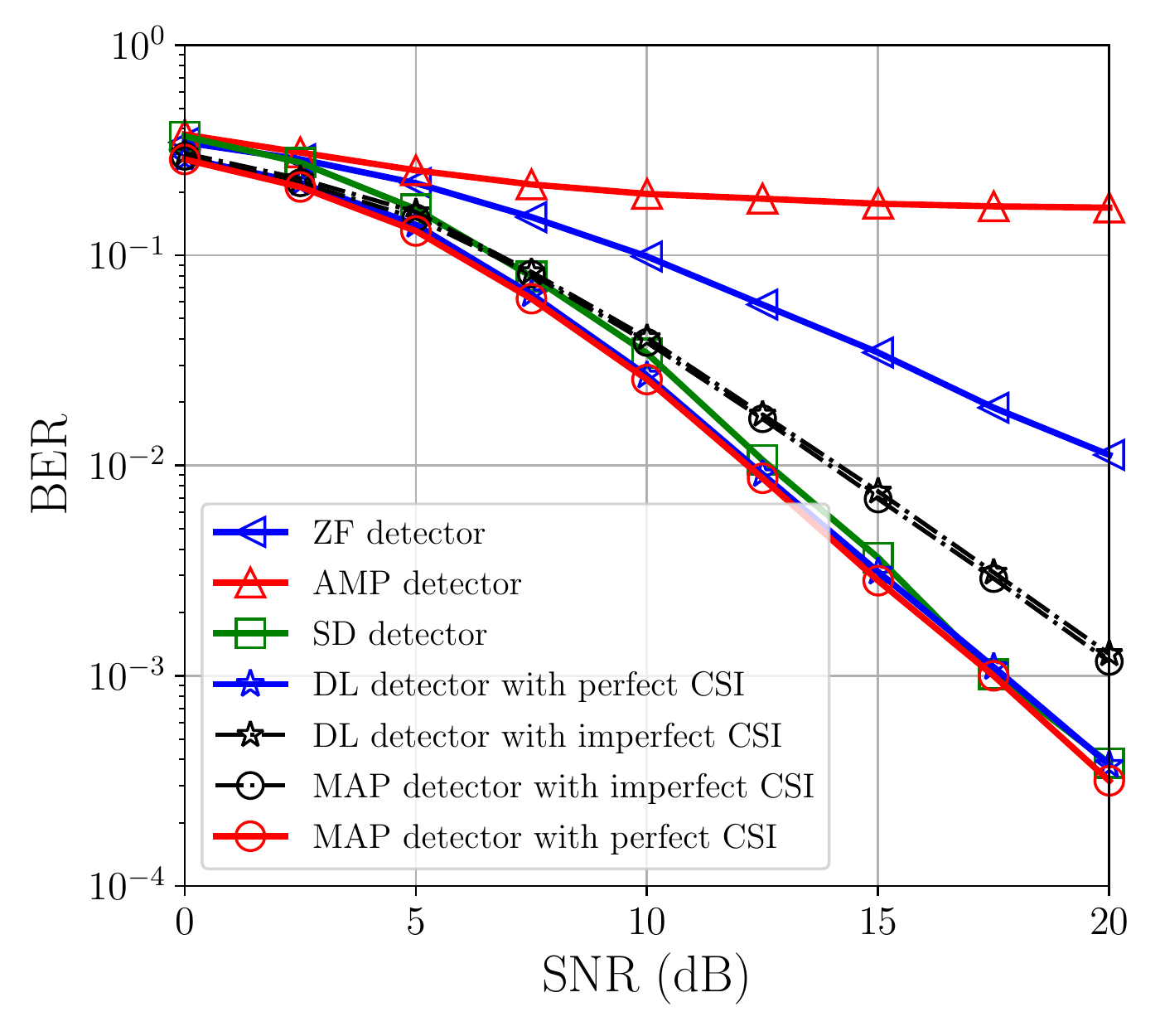}
				%\caption{Training data of Circle}
			\end{minipage}
		}
		%\hspace{1.4em}
		\subfigure[Correlated channel]{
			\begin{minipage}[t]{0.46\linewidth}
				\centering
				\includegraphics[width=\columnwidth]{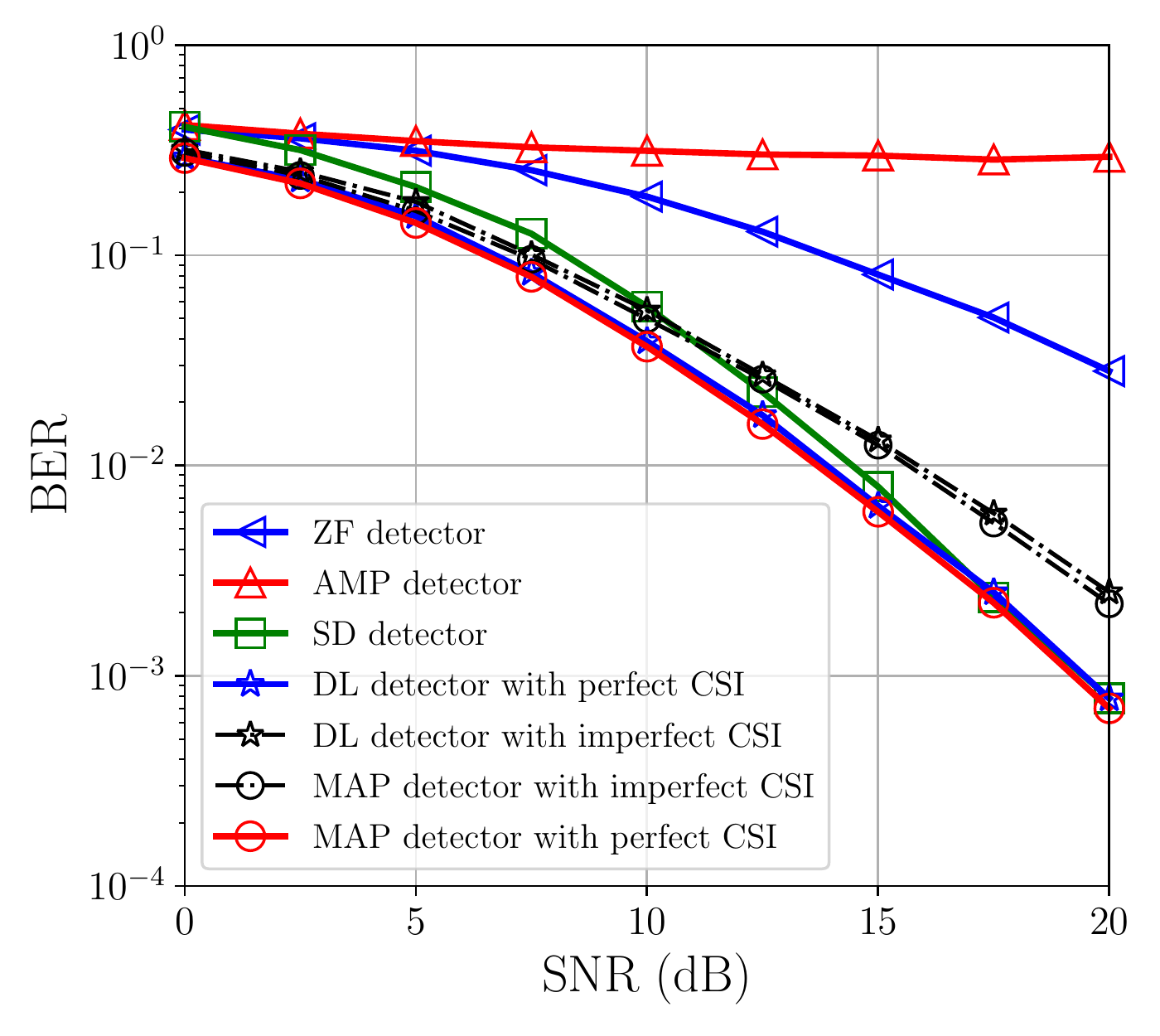}
				%\caption*{(b)}
			\end{minipage}
		}

		\caption{The BER performance of data-driven DL detector versus SNR compared to other MIMO detectors over the time-varying channel.}
		\label{figure6-1}
		\vskip -0.2in
	\end{figure*}

Fig.~\ref{figure6-1}(a) compares the BER performance of the model-based ZF, AMP, SD, MAP and the data-driven DL detectors versus SNR over the $2\times 2$ Gaussian channel. We assume that the ZF, AMP, and SD detectors have perfect CSI while both the DL and MAP detectors are evaluated under perfect and imperfect CSI, respectively. As illustrated in Fig.~\ref{figure5}(a), the data-driven DL detector manages to achieve the Bayes-optimal BER performance by incorporating the perfect $\mathbf{H}$ in time-varying channels and outperforms the other model-based detectors substantially. Fig.~\ref{figure5}(a) demonstrates that the data-driven DL detector can learn properly over time-varying channels. However, the BER performance of the data-driven DL detector is severely deteriorated by imperfect CSI in time-varying channels and can only approach to the MAP detector with imperfect CSI. 
	
	Fig~\ref{figure6-1}(b) compares the BER performance of the model-based ZF, AMP, SD, MAP and the data-driven DL detectors versus SNR over the $2\times 2$ correlated channel generated according to~\cite{shiu2000}. We assume that perfect CSI is available at the receiver. Fig~\ref{figure6-1}(b) shows that the BER performance of the data-driven DL detector coincides with that of the MAP detector and substantially outperforms other model-based algorithms, demonstrating its ability to achieve optimal accuracy in complex environments. Fig~\ref{figure5}(b) also indicates that the data-driven DL detector is model independent and manages to learn MAP comparable detection mapping over various channel models. Hence, the data-driven DL detector can keep the performance comparable with the MAP detector in various scenarios.

	\subsection{Nonlinear Systems}	
	In this subsection, we evaluate the BER performance of the data-driven DL detector under a nonlinear MIMO system. We will demonstrate that the data-driven DL detector is applicable to a broader range of scenarios than the traditional MIMO detection algorithms. Furthermore, we only compare the data-driven DL detector to the ZF and MAP detectors.
	
	Consider a MIMO system corrupted by the quantization error of ADC. We assume that each element of the channel output undergoes an entry-wise $B$ bit uniform quantizer $Q_{c}$. The channel input-output model is represented by~\eqref{nonlinear1} and can be rewritten as
	\begin{align}
		\mathbf{x}=Q_{c}(\mathbf{H}\mathbf{s}+\mathbf{n}).
	\end{align}
	Each real-valued input of $Q_{c}$ is mapped to one of $2^{B}$ bins, which are defined by the set of $2^{B}-1$ thresholds $[r_{1},r_{2},\ldots,r_{2^{B}-1}]$ such that $-\infty<r-{1}<r_{2}<\cdots<r_{2^{B}-1}<\infty$. Specifically, we define $r_{0}=-\infty$ and $r_{2^{B}}=\infty$. The threshold $r_{b}$ is given by
	\begin{align}
		r_{b}=\sqrt{B}(-2^{B-1}+b)2^{-B},\,\mathrm{for}\ b=1,\ldots,2^{B}-1,
	\end{align}
	where the quantization output of $Q_{c}$ is $r_{b}-\frac{\Delta}{2}$ when the input falls in the interval $(r_{b-1},r_{b}]$.\footnote{If $b=2^{B}$, the output of $Q_{c}$ is $\sqrt{B}(-2^{B-1}-2^{-1})2^{-B}$.}

\begin{figure}[!t]
		\centering
		\includegraphics[width=0.6\columnwidth]{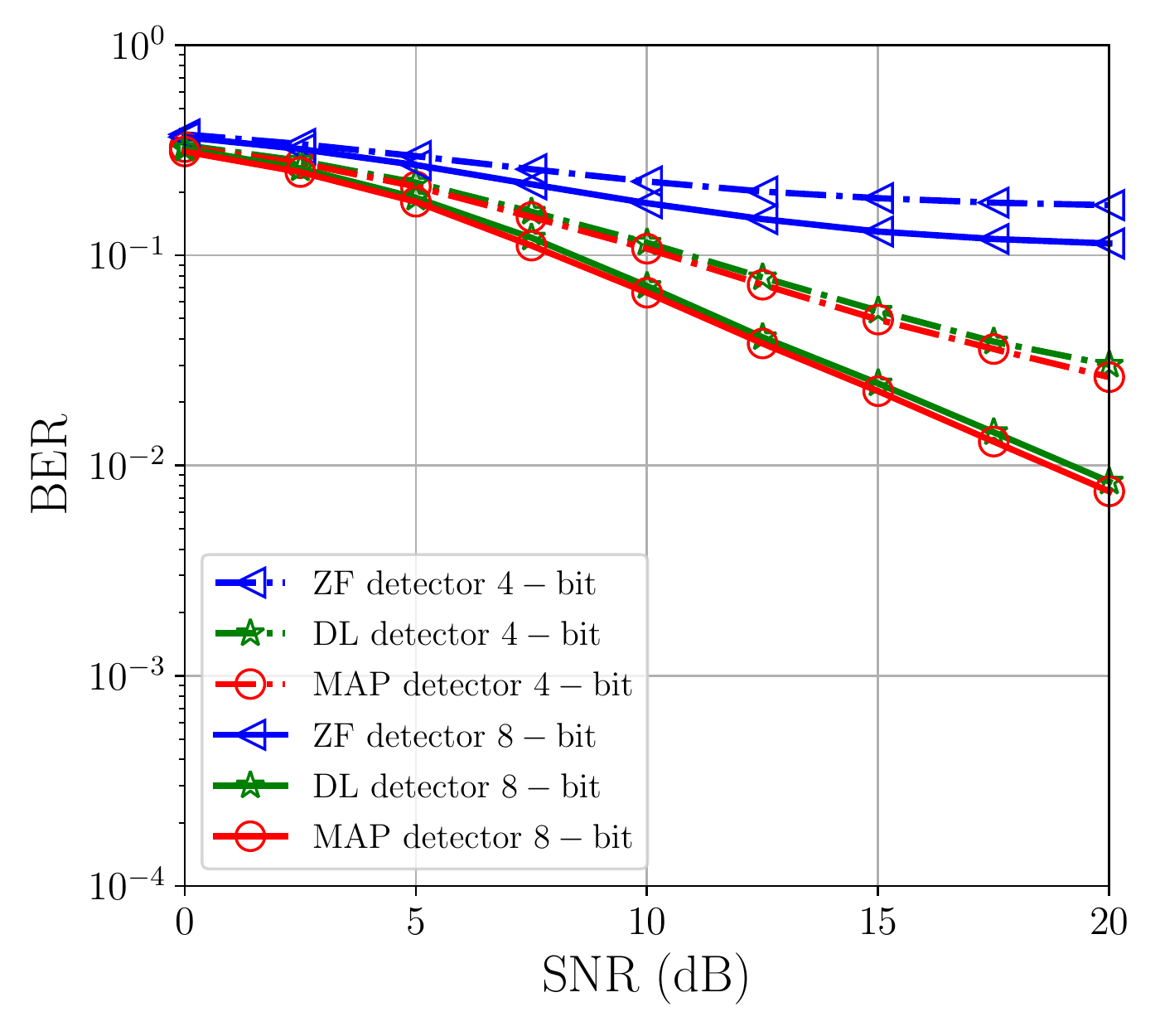}
		%where an .eps filename suffix will be assumed under latex, 
		% and a .pdf suffix will be assumed for pdflatex; or what has been declared
		% via \DeclareGraphicsExtensions.
		\caption{The BER performance of data-driven DL detector versus SNR compared to other MIMO detectors over the quantized time-varying channel.}
		\label{figure5}
	\end{figure}

	In Fig~\ref{figure5}, we compare the BER performance of the data-driven DL detector with the ZF and MAP detectors versus SNR over the quantized  $2\times 2$ time-varying Gaussian channel. The quantization bits are set to be $4$ and $8$, respectively. The perfect CSI is assumed to be available at the receiver. In Fig.~\ref{figure5}, the BER performance of the data-driven DL detector is close to that of the MAP detector under both $5$-bit and $10$-bit quantizers while the BER performance of the ZF detector degrades significantly. Hence, data-driven DL detector is also able to provide MAP comparable BER performance in quantized Gaussian channels. Fig.~\ref{figure5} also verifies model independence of the data-driven DL detector for nonlinear systems.

	\subsection{Model-driven DL Detector}

	\begin{figure}[!t]
		\centering
		\includegraphics[width=0.6\columnwidth]{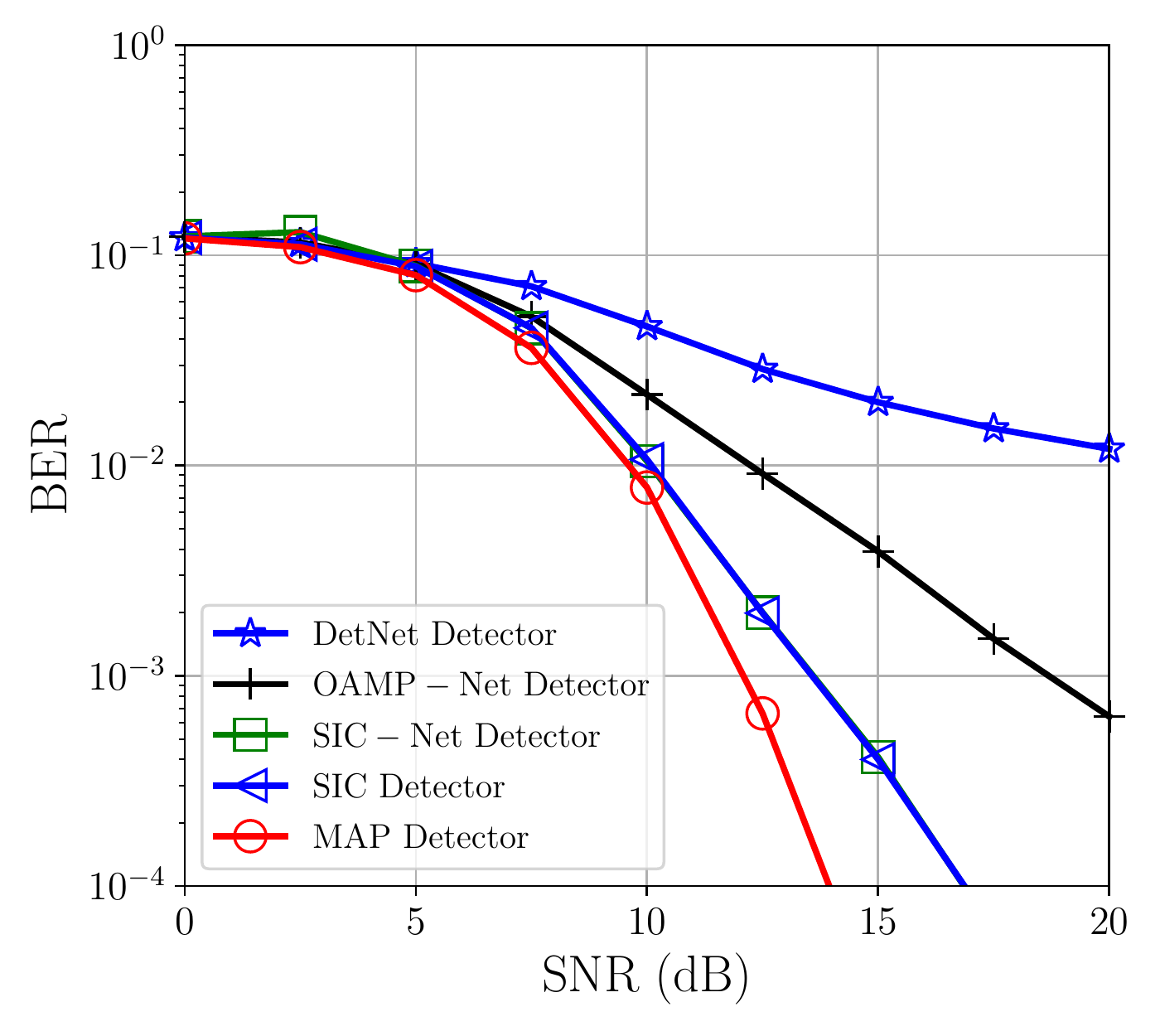}
		%where an .eps filename suffix will be assumed under latex, 
		% and a .pdf suffix will be assumed for pdflatex; or what has been declared
		% via \DeclareGraphicsExtensions.
		\caption{The BER performance of SIC-Net versus SNR compared to other MIMO detectors over the time-varying channel.}
		\label{figure6}
	\end{figure}

	In Fig.~\ref{figure6}, we evaluate the BER performance of the DetNet~\cite{Samuel2019Learning}, OAMP-Net~\cite{hedetec2020}, SIC~\cite{choi2000}, SIC-Net and MAP detectors over the $8\times 8$ time-varying Gaussian channel. We assume that perfect CSI is available at the receiver.  Fig.~\ref{figure6} shows that the BER performance of the SIC-Net detector is significantly better than those of the DetNet and OAMP-Net detectors but there still exits a gap between the SIC-Net and the MAP detectors, especially when SNR is high. Hence, the SIC-Net detector is suboptimal compared to the data-driven DL and MAP detectors. In Fig.~\ref{figure6}, the BER performance of the SIC detector is close to that of the SIC-Net detector. Theorem~\ref{theorem3} demonstrates that the SIC detector is equivalent to the optimal model-driven DL detector with the minimum $J_{u}(\mathbf{f}_{\boldsymbol{\vartheta}})$. Therefore, the performance of SIC-Net detector is determined by the SIC algorithm and the selection of detection algorithms is more important than trainable variables in improving the BER performance of the model-driven DL detectors.

	\section{Conclusions}\label{conclusion-sec}
	
	In this paper, we have made the first attempt on interpreting DL-based MIMO detection with two different deep architectures: DNN embedded data-driven DL detector and iterative model-driven DL detectors. We have showed that the data-driven DL detector can converge to the MAP detector in various scenarios under suitably configured structure and sufficiently large training set. Specifically, the data-driven DL detector is robust to CSI uncertainty in time-invariant channels and suffers from imperfect CSI in time-varying channels. Moreover, the data-driven DL detector is ineffective in large-sized MIMO systems due to its requirement on a large number of training samples. On the other hand, the model-driven DL detector successfully addresses this problem by exploiting model expert knowledge and achieves relatively good performance with only a small training set since its parameter space is with a small size. However, the model-driven DL detector is suboptimal compared to the MAP detector. The strengths and weaknesses of DL-based MIMO detection should be carefully balanced when deployed in different environments. 
	
	\section{Acknowledgement}
	Thanks for suggestions and comments from Dr. Shenglong Zhou of Imperial College London.
	
	\begin{appendices}
		\section{Proof for Theorem~\ref{theorem1}}\label{apex-a}
		From~\eqref{posterior}, we have
		\begin{align}\label{theo1-kl}
			J(p_{o})-J(p_{\boldsymbol{\theta}})&=\sum_{i=1}^{d_{l+1}}\int\ln\frac{ p_{o}(\tilde{\mathbf{u}}_{i}|\mathbf{x})}{ p_{\boldsymbol{\theta},i}(\mathbf{x})}f_{de}(\tilde{\mathbf{u}}_{i},\mathbf{x})d\mathbf{x}=\sum_{i=1}^{d_{l+1}}\int g_{\boldsymbol{\theta},i}(\mathbf{x})f_{de}(\tilde{\mathbf{u}}_{i},\mathbf{x})d\mathbf{x}
		\end{align}
		for any $\boldsymbol{\theta}\in\Theta_{R}$, where 
		\begin{align}\label{theo1-p}
			p_{\boldsymbol{\theta},i}(\mathbf{x})=\frac{\exp(f_{\boldsymbol{\theta},i}(\mathbf{x}))}{\sum_{j=1}^{d_{l+1}}\exp(f_{\boldsymbol{\theta},j}(\mathbf{x}))},
		\end{align}
		$f_{\boldsymbol{\theta},i}(\mathbf{x})$ is the $i$-th entry of $\mathbf{f}_{\boldsymbol{\theta}}(\mathbf{x})$, $g_{\boldsymbol{\theta},i}(\mathbf{x})=\ln\frac{ p_{o}(\tilde{\mathbf{u}}_{i}|\mathbf{x})}{ p_{\boldsymbol{\theta},i}(\mathbf{x})}$, and $f_{de}(\tilde{\mathbf{u}}_{i},\mathbf{x})$ is joint probability density function. 
		
		Since $J(p_{o})-J(p_{\boldsymbol{\theta}})\geq 0$, we obtain
		\begin{align}\label{theo1-inequal}
			J(p_{o})-J(p_{\boldsymbol{\theta}})&=|J(p_{o})-J(p_{\boldsymbol{\theta}})|\leq\sum_{i=1}^{d_{l+1}}\int |g_{\boldsymbol{\theta},i}(\mathbf{x})|f_{de}(\tilde{\mathbf{u}}_{i},\mathbf{x})d\mathbf{x}.
		\end{align}
		Substituting~\eqref{theo1-p} into $|g_{\boldsymbol{\theta},i}(\mathbf{x})|$ yields
		\begin{align}\label{theo1_epsilon2}
			|g_{\boldsymbol{\theta},i}(\mathbf{x})|=\bigg|\ln p_{o}(\tilde{\mathbf{u}}_{i}|\mathbf{x})-f_{\boldsymbol{\theta},i}(\mathbf{x})+\ln\bigg[\sum_{j=1}^{d_{l+1}}\exp(f_{\boldsymbol{\theta},j}(\mathbf{x}))\bigg]\bigg|.
		\end{align}	
		
		From~\eqref{network}, $f_{\boldsymbol{\theta},i}(x)$ is an $\mathbb{R}^{d_{0}}\rightarrow\mathbb{R}$ piecewise linear function. According to~\cite[Theorem 2.1]{arora2018understanding}, any $\mathbb{R}^{d_{0}}\rightarrow\mathbb{R}$ piecewise linear function can be represented by a ReLU DNN with no more than $\lceil\log_{2}(d_{0}+1)\rceil$ hidden layers. 
		
		Since $\ln p_{o}(\mathbf{u}|\mathbf{x})$ has finite 2-norm for every possible $\mathbf{u}$, each element of $\{\ln p_{o}(\tilde{\mathbf{u}}_{1}|\mathbf{x}),\ldots,\ln p_{o}(\tilde{\mathbf{u}}_{d_{l+1}}|\mathbf{x})\}$ can be approximated by a ReLU DNN with at most $\lceil\log_{2}(d_{0}+1)\rceil$ hidden layers~\cite{arora2018understanding,qiang2020channel}. We simply put these ReLU DNNs in parallel and combine their outputs together to compose a single ReLU DNN. As a result, there exits a DL detector with $\boldsymbol{\theta}_{\varepsilon}\in\Theta_{R}$ and at most 
		$\lceil\log_{2}(d_{0}+1)\rceil$ hidden layers
		such that
		\begin{align}\label{theo1_epsilon}
			|f_{\boldsymbol{\theta}_\varepsilon,i}(\mathbf{x})-\ln p_{o}(\tilde{\mathbf{u}}_{i}|\mathbf{x})|\leq\varepsilon
		\end{align}
		for any $\varepsilon>0$ and $i\in\{1,\ldots,d_{l+1}\}$.
		From~\eqref{theo1_epsilon} and~\eqref{theo1_epsilon2},  $|g_{\boldsymbol{\theta},i}(\mathbf{x})|$ satisfies
		\begin{align}\label{theo1_in}
			|g_{\boldsymbol{\theta}_\varepsilon,i}(\mathbf{x})|&\leq\big|\ln p_{o}(\tilde{\mathbf{u}}_{i}|\mathbf{x})-f_{\boldsymbol{\theta}_\varepsilon,i}(\mathbf{x})\big|+\bigg|\ln\bigg[\sum_{j=1}^{d_{l+1}}\exp(f_{\boldsymbol{\theta}_\varepsilon,j}(\mathbf{x}))\bigg]\bigg|\leq\varepsilon+\bigg|\ln\bigg[\sum_{j=1}^{d_{l+1}}\exp(f_{\boldsymbol{\theta}_\varepsilon,j}(\mathbf{x}))\bigg]\bigg|.
		\end{align}
		Specifically, $\sum_{j=1}^{d_{l+1}}\exp(f_{\boldsymbol{\theta}_\varepsilon,j}(\mathbf{x}))$ in~\eqref{theo1_in} is upper bounded by
		\begin{align}
			\sum_{j=1}^{d_{l+1}}\exp(f_{\boldsymbol{\theta}_\varepsilon,j}(\mathbf{x}))\leq e^{\varepsilon}\sum_{j=1}^{d_{l+1}}p_{o}(\tilde{\mathbf{u}}_{j}|\mathbf{x})=e^{\varepsilon}
		\end{align}
		and is lower bounded by
		\begin{align}
			\sum_{j=1}^{d_{l+1}}\exp(f_{\boldsymbol{\theta}_\varepsilon,j}(\mathbf{x}))\geq e^{-\varepsilon}.
		\end{align}
		Then, 
		\begin{align}
			\bigg|\ln
			\bigg[\sum_{j=1}^{d_{l+1}}\exp(f_{\boldsymbol{\theta}_\varepsilon,j}(\mathbf{x}))\bigg]\bigg|\leq \varepsilon
		\end{align}
		and
		\begin{align}\label{theo1_g}
			|g_{\boldsymbol{\theta}_\varepsilon,i}(\mathbf{x})|\leq 2\varepsilon
		\end{align}
		hold. Substituting~\eqref{theo1_g} into~\eqref{theo1-inequal} yields
		\begin{align}
			J(p_{o})-J(p_{\boldsymbol{\theta}_\varepsilon})\leq 2\varepsilon\sum_{i=1}^{d_{l+1}}\int f_{de}(\tilde{\mathbf{u}}_{i},\mathbf{x})d\mathbf{x}=2\varepsilon.
		\end{align}
		
		From~\eqref{mini1}, $p_{\boldsymbol{\theta}_{o}}(\mathbf{u}|\mathbf{x})$ has the lowest  KL information for all $\boldsymbol{\theta}\in\Theta_{R}$ and we have
		\begin{align}\label{theo1-2}
			J(p_{o})-J(p_{\boldsymbol{\theta}_{o}})\leq J(p_{o})-J(p_{\boldsymbol{\theta}_\varepsilon})\leq 2\varepsilon.
		\end{align}
		It is then easy to derive~\eqref{theo1} from~\eqref{theo1-2} since $\varepsilon$ is an arbitrary positive value, which completes the proof.
		
		\section{Proof for Lemma~\ref{lemma-1}}\label{apex-b}
		According to Symmetrization Lemma in~\cite[Chapter II.3]{pollard2012convergence}, the inequality in~\eqref{lemma1} holds if
		\begin{align}\label{lemma1-1}
			\mathbf{P}(|J_{\mathcal{Z}}(p_{\boldsymbol{\theta}})-J(p_{\boldsymbol{\theta}})|>\frac{\varepsilon}{2})\leq\frac{1}{2}
		\end{align}
		for all $\boldsymbol{\theta}\in\Theta_{R}$. Let $\sigma^{2}(p_{\boldsymbol{\theta}})$ be the variance of $\ln p_{\boldsymbol{\theta}}(\mathbf{u}|\mathbf{x})$. Using Chebyshev's inequality~\cite{mendenhall2012introduction} yields
		\begin{align}\label{chebshev}
			\mathbf{P}\big(|J_{\mathcal{Z}}(p_{\boldsymbol{\theta}})-J(p_{\boldsymbol{\theta}})|\geq\frac{\varepsilon}{2}\big)\leq\frac{4\sigma^{2}(p_{\boldsymbol{\theta}})}{|\mathcal{Z}|\varepsilon^{2}}
		\end{align}
		for all $\boldsymbol{\theta}\in\Theta_{R}$. Specifically, $\sigma^{2}(p_{\boldsymbol{\theta}})$ satisfies
		\begin{align}\label{app-theo4-3}
			\sigma^{2}(p_{\boldsymbol{\theta}})&=\mathbb{E}\big\{[\ln p_{\boldsymbol{\theta}}(\mathbf{u}|\mathbf{x})]^{2}\big\}-J(p_{\boldsymbol{\theta}})^{2}\leq\mathbb{E}\big\{[\ln p_{\boldsymbol{\theta}}(\mathbf{u}|\mathbf{x})]^{2}\big\}\nonumber\\
			&\leq\sum_{i=1}^{d_{l+1}}\int\big[|f_{\boldsymbol{\theta},i}(\mathbf{x})|+\big|\ln(\sum_{j=1}^{d_{l+1}}\exp(f_{\boldsymbol{\theta},j}(\mathbf{x})))\big|\big]^{2}f_{de}(\tilde{\mathbf{u}}_{i},\mathbf{x})d\mathbf{x}.
		\end{align}
		
		Assume that $\mathcal{X}$ follows the partition in~\eqref{partition} and use the fact~\eqref{network}. The triangle inequality assures that
		\begin{align}\label{lemma1-3}
			|f_{\boldsymbol{\theta},i}(\mathbf{x})|&=\|\mathbf{w}_{\mathcal{X}_{k},i}\mathbf{x}+b_{\mathcal{X}_{k},i}\|_{2}\leq\|\mathbf{w}_{\mathcal{X}_{k},i}\|_{2}\|\mathbf{x}\|_{2}+|b_{\mathcal{X}_{k},i}|
		\end{align} 
		for $\mathbf{x}\in\mathcal{X}_{k}$, where $\mathbf{w}_{\mathcal{X}_{k},i}$ and $b_{\mathcal{X}_{k},i}$ are the $i$-th row and the $i$-th entry of $\mathbf{W}_{\mathcal{X}_{k}}$ and $\mathbf{b}_{\mathcal{X}_{k}}$, respectively. 
		
		From~\eqref{hyperplane} and~\eqref{network}, $\|\mathbf{w}_{\mathcal{X}_{k},i}\|_{2}$ and $\|b_{\mathcal{X}_{k},i}\|_{2}$ in~\eqref{lemma1-3} are upper bounded by
		\begin{align}\label{weight}
			\|\mathbf{w}_{\mathcal{X}_{k},i}\|_{2}&=\big\|\tilde{\mathbf{w}}_{l,i}\prod_{j=0}^{l-1}\tilde{\mathbf{W}}_{j}\big\|_{2}=\big\|\mathbf{w}_{l,i}\mathbf{\Lambda}_{l}\prod_{j=0}^{l-1}\mathbf{W}_{j}\mathbf{\Lambda}_{j}\big\|_{2}\leq\|\mathbf{w}_{l,i}\mathbf{\Lambda}_{l}\|_{2}\prod_{j=0}^{l-1} \big\|\mathbf{W}_{j}\mathbf{\Lambda}_{j}\big\|_{2}\leq\|\mathbf{w}_{l,i}\|_{2}\prod_{j=0}^{l-1} \big\|\mathbf{W}_{j}\big\|_{2}
		\end{align}
		and
		\begin{align}\label{bias}
			|b_{\mathcal{X}_{k},i}|&=\Big|\sum_{j=0}^{l-1}\tilde{\mathbf{w}}_{l,i}\bigg(\prod_{q=0}^{j-1}\tilde{\mathbf{W}}_{l-1-q}\bigg)\mathbf{b}_{l-1-j}+b_{l,i}\Big|\leq\sum_{j=0}^{l-1}\|\tilde{\mathbf{w}}_{l,i}\|_{2}\big\|\prod_{q=0}^{j-1}\tilde{\mathbf{W}}_{l-1-q}\big\|_{2}\|\mathbf{b}_{l-1-j}\|_{2}+|b_{l,i}|\nonumber\\
			&\leq\sum_{j=0}^{l-1}\|\mathbf{w}_{l,i}\|_{2}\bigg(\prod_{q=0}^{j-1}\|\mathbf{W}_{l-1-q}\|_{2}\bigg)\|\mathbf{b}_{l-1-j}\|_{2}+|b_{l,i}|,
		\end{align}
		respectively, where $\tilde{\mathbf{w}}_{l,i}$ and $\mathbf{w}_{l,i}$ are the $i$-th rows of $\tilde{\mathbf{W}}_{l}$ and $\mathbf{W}_{l}$ and $b_{l,i}$ is the $i$-th entry of $\mathbf{b}_{l}$. Since $\|\mathbf{W}_{j}\|_{2}\leq R\|\mathbf{d}\|_{\infty}=\alpha$, $\|\mathbf{b}_{j}\|_{2}\leq\sqrt{R}\|\mathbf{d}\|_{\infty}\leq\alpha$, $\|\mathbf{w}_{l,i}\|_{2}\leq\sqrt{R}\|\mathbf{d}\|_{\infty}\leq\alpha$, and $|b_{l,i}|\leq\|\mathbf{d}\|_{\infty}\leq\alpha$ for $j\in\{0,1,\ldots,l-1\}$,~\eqref{weight} and~\eqref{bias} can be further bounded by
		\begin{align}\label{lemma1-weight1}
			\|\mathbf{w}_{\mathcal{X}_{k},i}\|_{2}\leq\alpha^{l+1}
		\end{align}
		and
		\begin{align}\label{lemma1-bias1}
			&|b_{\mathcal{X}_{k},i}|\leq\bigg(\sum_{i=0}^{l-1}\alpha^{i+1}\bigg)\alpha+\alpha=\frac{\alpha^{l+2}-\alpha}{\alpha-1}\leq\beta(\alpha^{l+1}-1),
		\end{align}
		respectively. Substituting~\eqref{lemma1-3},~\eqref{lemma1-weight1}, and~\eqref{lemma1-bias1} into~\eqref{lemma1-3} yields
		\begin{align}\label{lemma1-4}
			|f_{\boldsymbol{\theta},i}(\mathbf{x})|&\leq \alpha^{l+1}(\|\mathbf{x}\|_{2}+\beta)-\beta.
		\end{align}
		Using~\eqref{lemma1-4},  $\big|\ln[\sum_{i=1}^{d_{l+1}}\exp(f_{\boldsymbol{\theta},i}(\mathbf{x}))]\big|$ in~\eqref{app-theo4-3} is upper bounded by
		\begin{align}\label{lemma1-5}
			\bigg|\ln\bigg[\sum_{i=1}^{d_{l+1}}\exp(f_{\boldsymbol{\theta},i}(\mathbf{x}))\bigg]\bigg|\leq \ln d_{l+1}\big[\alpha^{l+1}(\|\mathbf{x}\|_{2}+\beta)-\beta\big].
		\end{align}	
		Combining~\eqref{lemma1-4} and~\eqref{lemma1-5}, we have
		\begin{align}\label{lemma1-2}
			\sigma^{2}(p_{\boldsymbol{\theta}})&\leq\sum_{i=1}^{d_{l+1}}\int\big[(\ln d_{l+1}+1)(\alpha^{l+1}(\|\mathbf{x}\|_{2}+\beta)-\beta)\big]^{2}f_{de}(\tilde{\mathbf{u}}_{i},\mathbf{x})d\mathbf{x}
			\nonumber\\
			&=\mathbb{E}\big\{\big[(\ln d_{l+1}+1)(\alpha^{l+1}(\|\mathbf{x}\|_{2}+\beta)-\beta)\big]^{2}\big\}=\nu.
		\end{align}

		Replacing $\sigma^{2}(p_{\boldsymbol{\theta}})$ in~\eqref{chebshev} by its bound in~\eqref{lemma1-2} and letting $|\mathcal{Z}|\geq 4\nu/\varepsilon^{2}$, we obtain the inequalities in~\eqref{lemma1} and~\eqref{lemma1-1},
		which completes the proof.
		
		\section{Proof for Lemma~\ref{lemma2}}\label{apex-c}
		Let $\mathcal{Z}_{i}$ denote the index set of training samples in~$\mathcal{Z}$ where
		\begin{align}
			\mathbf{u}_{j}=\tilde{\mathbf{u}}_{i},\ \forall j\in\mathcal{Z}_{i}
		\end{align}
		for $i\in\{1,\ldots,d_{l+1}\}$. Since samples in $\mathcal{Z}_{i}$ are generated according to $\tilde{\mathbf{u}}_{i}$ and $(\tilde{\mathbf{u}}_{1},\ldots,\tilde{\mathbf{u}}_{d_{l+1}})$ is an i.i.d. multinomial random variable, $(|\mathcal{Z}_{1}|,\ldots,|\mathcal{Z}_{d_{l+1}}|)$ is also an i.i.d. multinomial random variable with probability $(p(\tilde{\mathbf{u}}_{1}),\ldots,p(\tilde{\mathbf{u}}_{d_{l+1}}))$  and $\sum_{i=1}^{d_{l+1}}|\mathcal{Z}_{i}|=|\mathcal{Z}|$.	Then, $|J_{\mathcal{Z}}(p_{\boldsymbol{\theta}})-J_{\mathcal{Z}}(p_{\boldsymbol{\lambda}})|$ is upper bounded by 
		\begin{align}\label{lemma2-1}
			&|J_{\mathcal{Z}}(p_{\boldsymbol{\theta}})-J_{\mathcal{Z}}(p_{\boldsymbol{\lambda}})|\leq\frac{1}{|\mathcal{Z}|}\sum_{i=1}^{d_{l+1}}\sum_{j\in\mathcal{Z}_{i}}|\ln p_{\boldsymbol{\theta}}(\mathbf{u}_{j}|\mathbf{x}_{j})-\ln p_{\boldsymbol{\lambda}}(\mathbf{u}_{j}|\mathbf{x}_{j})|\nonumber\\
			&\leq\frac{1}{|\mathcal{Z}|}\sum_{i=1}^{d_{l+1}}\sum_{j\in\mathcal{Z}_{i}}\bigg|f_{\boldsymbol{\theta},i}(\mathbf{x}_{j})-f_{\boldsymbol{\lambda},i}(\mathbf{x}_{j})-\ln\bigg[\sum_{i=1}^{d_{l+1}}\exp(f_{\boldsymbol{\theta},i}(\mathbf{x}_{j}))\bigg]+\ln\bigg[\sum_{i=1}^{d_{l+1}}\exp(f_{\boldsymbol{\lambda},i}(\mathbf{x}_{j}))\bigg]\bigg|\nonumber\\
			&\leq \frac{1}{|\mathcal{Z}|}\sum_{i=1}^{d_{l+1}}\sum_{j\in\mathcal{Z}_{i}}\big|f_{\boldsymbol{\theta},i}(\mathbf{x}_{j})-f_{\boldsymbol{\lambda},i}(\mathbf{x}_{j})\big|+\bigg|\ln\bigg[\sum_{i=1}^{d_{l+1}}\exp(f_{\boldsymbol{\theta},i}(\mathbf{x}_{j}))\bigg]-\ln\bigg[\sum_{i=1}^{d_{l+1}}\exp(f_{\boldsymbol{\lambda},i}(\mathbf{x}_{j}))\bigg]\bigg|.
		\end{align}
		
		Denote $\mathbf{V}_{i}$ and $\boldsymbol{\upsilon}_{i}$ as the weight and the bias of the $i$-th layer of $\mathcal{D}$ corresponding to $\boldsymbol{\lambda}$ for $i\in\{0,\ldots,l\}$. Let
		\begin{align}
			\boldsymbol{\theta}_{j}=(\mathrm{vec}(\mathbf{W}_{0}),\mathbf{b}_{0},\ldots,\mathrm{vec}(\mathbf{W}_{j-1}),\mathbf{b}_{j-1})
		\end{align}
		and
		\begin{align}
			\boldsymbol{\lambda}_{j}=(\mathrm{vec}(\mathbf{V}_{0}),\boldsymbol{\upsilon}_{0},\ldots,\mathrm{vec}(\mathbf{V}_{j-1}),\boldsymbol{\upsilon}_{j-1}),
		\end{align}
		be the partial parameters up to the $j$-th layer for $j\in\{1,\ldots,l+1\}$. Hence, the corresponding partial network outputs are denoted by $\mathbf{f}_{\boldsymbol{\theta}_{j}}(\mathbf{x})$ and $\mathbf{f}_{\boldsymbol{\lambda}_{j}}(\mathbf{x})$, respectively. 
		
		Let $e_{j}=\|\mathbf{f}_{\boldsymbol{\theta}_{j}}(\mathbf{x})-\mathbf{f}_{\boldsymbol{\lambda}_{j}}(\mathbf{x})\|_{2}$
		be the partial error for $j\in\{1,\ldots,l+1\}$ and $e_{l+1}$ represents $\big|f_{\boldsymbol{\theta},i}(\mathbf{x}_{m})-f_{\boldsymbol{\lambda},i}(\mathbf{x}_{m})\big|$ in~\eqref{lemma2-1}. According to~\cite[Lemma~2]{qiang2020channel} and Lemma~\ref{lemma-1}, the upper bound on $e_{j+1}$ is given by
		\begin{align}\label{lemma2-2}
			e_{j+1}\leq r\|\mathbf{d}\|_{\infty}\big[(3\alpha)^{j}(\|\mathbf{x}\|_{2}+1)+\sum_{q=0}^{j-1}(3\alpha)^{q}(y_{j-q}+1)\big]
		\end{align}
		for $j\in\{1,\ldots,l\}$, where $r=\|\boldsymbol{\theta}-\boldsymbol{\lambda}\|_{\infty}\leq 2R$ and
		$y_{j}=\alpha^{j}(\|\mathbf{x}\|_{2}+\beta)-\beta$ is the upper bound on $\|\mathbf{f}_{\boldsymbol{\theta}_{j}}(\mathbf{x})\|_{2}$ and $\|\mathbf{f}_{\boldsymbol{\lambda}_{j}}(\mathbf{x})\|_{2}$.
		
		From~\eqref{lemma2-2}, we know $|f_{\boldsymbol{\theta},i}(\mathbf{x})-f_{\boldsymbol{\lambda},i}(\mathbf{x})|$ is upper bounded by
		\begin{align}\label{lemma2-4}
			|f_{\boldsymbol{\theta},i}(\mathbf{x})-f_{\boldsymbol{\lambda},i}(\mathbf{x})|&\leq r\|\mathbf{d}\|_{\infty}\bigg[\sum_{q=0}^{l} 3^{q}\alpha^{l}\|\mathbf{x}\|_{2}+(3\alpha)^{l}+\sum_{q=0}^{l-1}\big[3^{q}\alpha^{l}\beta-(3\alpha)^{q}(\beta-1)\big]\bigg]\nonumber\\
			&= r\|\mathbf{d}\|_{\infty}\bigg[ \frac{3^{l+1}-1}{2}\alpha^{l}\|\mathbf{x}\|_{2}+(3\alpha)^{l}+\frac{3^{l}-1}{2}\alpha^{l}\beta -\frac{(3\alpha)^{l}-1}{3\alpha-1}(\beta-1)\bigg]\nonumber\\
			&\leq \frac{3}{2}r\|\mathbf{d}\|_{\infty}(3\alpha)^{l}(\|\mathbf{x}\|_{2}+\beta)=\xi r,
		\end{align}
		where $\xi =\frac{3}{2}\|\mathbf{d}\|_{\infty}(3\alpha)^{l}(\|\mathbf{x}\|_{2}+\beta)$.
		From~\eqref{lemma2-4}, the upper bound on $\ln(\sum_{i=1}^{d_{l+1}}\exp(f_{\boldsymbol{\theta},i}(\mathbf{x})))$ can be expressed as
		\begin{align}
			\ln\bigg[\sum_{i=1}^{d_{l+1}}\exp(f_{\boldsymbol{\theta},i}(\mathbf{x}))\bigg]&\leq\ln\bigg[\sum_{i=1}^{d_{l+1}}\exp(f_{\boldsymbol{\lambda},i}(\mathbf{x})+\xi r)\bigg]=\xi r+\ln\bigg[\sum_{i=1}^{d_{l+1}}\exp(f_{\boldsymbol{\lambda},i}(\mathbf{x})\bigg].
		\end{align}
		Similarly,  $\ln\big[\sum_{i=1}^{d_{l+1}}\exp(f_{\boldsymbol{\theta},i}(\mathbf{x}))\big]$ is lower bounded by
		\begin{align}
			\ln\bigg[\sum_{i=1}^{d_{l+1}}\exp(f_{\boldsymbol{\theta},i}(\mathbf{x}))\bigg]\geq\ln\bigg[\sum_{i=1}^{d_{l+1}}\exp(f_{\boldsymbol{\lambda},i}(\mathbf{x}))\bigg]-\xi r.
		\end{align}
		Therefore, we have
		\begin{align}\label{lemma2-6}
			\bigg|\ln\bigg[\sum_{i=1}^{d_{l+1}}\exp(f_{\boldsymbol{\theta},i}(\mathbf{x}))\bigg]-\ln\bigg[\sum_{i=1}^{d_{l+1}}\exp(f_{\boldsymbol{\lambda},i}(\mathbf{x}))\bigg]\bigg|\leq\xi r.
		\end{align}
		
		Using~\eqref{lemma2-4} and~\eqref{lemma2-6}, we derive the upper bound on $|J_{\mathcal{Z}}(p_{\boldsymbol{\theta}})-J_{\mathcal{Z}}(p_{\boldsymbol{\lambda}})|$ as
		\begin{align}
			|J_{\mathcal{Z}}(p_{\boldsymbol{\theta}})-J_{\mathcal{Z}}(p_{\boldsymbol{\lambda}})|&\leq \frac{1}{|\mathcal{Z}|}\sum_{m=1}^{|\mathcal{Z}|}3r\|\mathbf{d}\|_{\infty}(3\alpha)^{l}(\|\mathbf{x}_{m}\|_{2}+\beta)\leq 3^{l+1}\|\mathbf{d}\|_{\infty}\alpha^{l}(\delta+\beta)\|\boldsymbol{\theta}-\boldsymbol{\lambda}\|_{\infty},
		\end{align}
		which completes the proof.
		\section{Proof for Theorem~\ref{theorem2}}\label{apex-d}
		From~\eqref{mini1} and~\eqref{min2}, we can bound $J(p_{\boldsymbol{\theta}_{o}})-J(p_{\boldsymbol{\theta}_{\mathcal{Z}}})$ by
		\begin{align}\label{theorem2-1}
			0&\leq J(p_{\boldsymbol{\theta}_{o}})-J(p_{\boldsymbol{\theta}_{\mathcal{Z}}})=[J(p_{\boldsymbol{\theta}_{o}})-J(p_{\boldsymbol{\theta}_{\mathcal{Z}}})]-[J_{\mathcal{Z}}(p_{\boldsymbol{\theta}_{o}})-J_{\mathcal{Z}}(p_{\boldsymbol{\theta}_{\mathcal{Z}}})]+[J_{\mathcal{Z}}(p_{\boldsymbol{\theta}_{o}})-J_{\mathcal{Z}}(p_{\boldsymbol{\theta}_{\mathcal{Z}}})]\nonumber\\
			&\leq [J(p_{\boldsymbol{\theta}_{o}})-J(p_{\boldsymbol{\theta}_{\mathcal{Z}}})]-[J_{\mathcal{Z}}(p_{\boldsymbol{\theta}_{o}})-J_{\mathcal{Z}}(p_{\boldsymbol{\theta}_{\mathcal{Z}}})]\leq 2\sup_{\boldsymbol{\theta}\in\Theta_{R}}|J_{\mathcal{Z}}(p_{\boldsymbol{\theta}})-J(p_{\boldsymbol{\theta}})|.
		\end{align}
		According to~\eqref{theorem2-1} and Lemma~\ref{lemma-1}, we know the following inequalities
		\begin{align}\label{theorem2-7}
			\mathbf{P}([J(p_{\boldsymbol{\theta}_{o}})-J(p_{\boldsymbol{\theta}_{\mathcal{Z}}})]>\varepsilon)&\leq	\mathbf{P}\big(\sup_{\boldsymbol{\theta}\in\Theta_{R}}|J_{\mathcal{Z}}(p_{\boldsymbol{\theta}})-J(p_{\boldsymbol{\theta}})|>\frac{\varepsilon}{2}\big)\leq 4\mathbf{P}\big(\sup_{\boldsymbol{\theta}\in\Theta_{R}}|J_{\mathcal{Z}}^{\circ}(p_{\boldsymbol{\theta}})|>\frac{\varepsilon}{8}\big),
		\end{align}
		holds if $|\mathcal{Z}|\geq 16\nu/\varepsilon^{2}$.
		
		Assume that $\mathcal{Z}$ is fixed with $\frac{1}{|\mathcal{Z}|}\sum_{m=1}^{|\mathcal{Z}|}\|\mathbf{x}_{m}\|_{2}^{2}\leq\delta^{2}$. Let $C=C(\varepsilon/16,\Theta_{R})$ and choose a collection of functions $\mathbf{p}_{1}(\mathbf{x}),\ldots,\mathbf{p}_{C}(\mathbf{x})\in\mathcal{D}$ such that 
		\begin{align}\label{theorem2-2}
			|J_{\mathcal{Z}}(p_{\boldsymbol{\theta}})-J_{\mathcal{Z}}(p_{j})|\leq\frac{\varepsilon}{16},\ \forall j\in\{1,\ldots,C\},
		\end{align}  
		for any $p_{\boldsymbol{\theta}}(\mathbf{x})\in\mathcal{D}$ and $\boldsymbol{\theta}\in\Theta_{R}$. According to~\cite[Theorem~2]{qiang2020channel}, the following inequality
		\begin{align}\label{theorem2-3}
			\mathbf{P}\Big(\sup_{\boldsymbol{\theta}\in\Theta_{R}}|J_{\mathcal{Z}}^{\circ}(\mathbf{f}_{\boldsymbol{\theta}})|>\frac{\varepsilon}{8}|\mathcal{Z}\Big)\leq\sum_{j=1}^{C} 2\mathrm{exp}\bigg[-2(\frac{|\mathcal{Z}|}{16}\varepsilon)^{2}/\sum_{m=1}^{|\mathcal{Z}|}(2\ln p_{j}(\mathbf{u}_{m}|\mathbf{x}_{m}))^{2}\bigg]
		\end{align} 
		holds. From~\eqref{app-theo4-3} and~\eqref{lemma1-2}, we know $\sum_{m=1}^{|\mathcal{Z}|}(\ln p(\mathbf{u}_{m}|\mathbf{x}_{m}))^{2}$ is upper bounded by 
		\begin{align}\label{theorem2-5}
			\sum_{m=1}^{|\mathcal{Z}|}(\ln p(\mathbf{u}_{m}|\mathbf{x}_{m}))^{2}&\leq \sum_{m=1}^{|\mathcal{Z}|}\big[(\ln d_{l+1}+1)(\alpha^{l+1}(\|\mathbf{x}_{m}\|_{2}+\beta)-\beta)\big]^{2} \nonumber\\
			&\leq |\mathcal{Z}|\big[(\ln d_{l+1}+1)(\alpha^{l+1}\delta+\beta)-\beta)\big]^{2}=|\mathcal{Z}|\delta_{\mathrm{1}}.
		\end{align}

		Replacing $\sum_{m=1}^{|\mathcal{Z}|}(\ln p(\mathbf{u}_{m}|\mathbf{x}_{m}))^{2}$ in~\eqref{theorem2-3} by the upper bound in~\eqref{theorem2-5}  yields
		\begin{align}\label{theorem2-6}
			\mathbf{P}\Big(\sup_{\boldsymbol{\theta}\in\Theta_{R}}|J_{\mathcal{Z}}^{\circ}(\mathbf{f}_{\boldsymbol{\theta}})|>\frac{\varepsilon}{8}|\mathcal{Z}\Big)\leq 2\mathrm{exp}\Big(\ln C-\frac{|\mathcal{Z}|\varepsilon^{2}}{512\delta_{\mathrm{1}}}\Big).
		\end{align}
		
		According to~\eqref{covering}, $\ln C$ is upper bounded by
		\begin{align}
			&\ln C =\ln C(\varepsilon/16,\Theta_{R})\leq d_{s}\ln\Big[\frac{3^{l+1}2^{6}\alpha^{l}(\delta+\beta)}{\varepsilon}\Big]\leq d_{s}\ln\frac{\delta_{2}}{\varepsilon}.
		\end{align}
		
		If $|\mathcal{Z}|\geq (1024\delta_{1}d_{s}\ln\frac{\delta_{2}}{\varepsilon})/\varepsilon^{2}$,
		then we have $
		\ln C\leq |\mathcal{Z}|\varepsilon^{2}/(1024\delta_{1})$
		and
		\begin{align}\label{theorem2-11}
			\mathbf{P}\Big(\sup_{\boldsymbol{\theta}\in\Theta_{R}}|J_{\mathcal{Z}}^{\circ}(p_{\boldsymbol{\theta}})|>\frac{\varepsilon}{8}|\mathcal{Z}\Big)\leq 2\mathrm{exp}\Big(-\frac{|\mathcal{Z}|\varepsilon^{2}}{1024\delta_{\mathrm{1}}}\Big).
		\end{align}

		Integrating out $\mathbf{P}(\sup_{\boldsymbol{\theta}\in\Theta_{R}}|J_{\mathcal{Z}}^{\circ}(p_{\boldsymbol{\theta}})|>\frac{\varepsilon}{8}|\mathcal{Z})$ over $\mathcal{Z}$ in~\eqref{theorem2-11} produces
		\begin{align}\label{theorem2-8}
			&\mathbf{P}\Big(\sup_{\boldsymbol{\theta}\in\Theta_{R}}|J_{\mathcal{Z}}^{\circ}(\mathbf{f}_{\boldsymbol{\theta}})|>\frac{\varepsilon}{8}\Big)\leq 2\mathrm{exp}
			\Big(-\frac{|\mathcal{Z}|\varepsilon^{2}}{1024\delta_{\mathrm{1}}}\Big)+\mathbf{P}_{\mathcal{Z}},
		\end{align}
		where
		\begin{align}
			\mathbf{P}_{\mathcal{Z}}=\mathbf{P}\Big(\frac{1}{|\mathcal{Z}|}\sum_{m=1}^{|\mathcal{Z}|}\|\mathbf{x}_{m}\|_{2}^{2}\geq\delta^{2}\Big).
		\end{align}
		
		If $\delta^{2}>\mu$, then Chebyshev's inequality~\cite{mendenhall2012introduction} assures that
		\begin{align} \mathbf{P}_{\mathcal{Z}}&\leq\mathbf{P}\big[|\frac{1}{|\mathcal{Z}|}\sum_{m=1}^{|\mathcal{Z}|}\|\mathbf{x}_{m}\|_{2}^{2}-\mu|\geq(\delta^{2}-\mu)\big]\leq\frac{\sigma^{2}}{|\mathcal{Z}|(\delta^{2}-\mu)^{2}}.
		\end{align}
		Combining~\eqref{theorem2-7} and~\eqref{theorem2-8}, we obtain
		\begin{align}\label{theorem2-9}
			\mathbf{P}([J(p_{\boldsymbol{\theta}_{o}})-J(p_{\boldsymbol{\theta}_{\mathcal{Z}}})]>\varepsilon)\leq 8\mathrm{exp}\Big(-\frac{|\mathcal{Z}|\varepsilon^{2}}{1024\delta_{\mathrm{1}}}\Big)+\frac{4\sigma^{2}}{|\mathcal{Z}|(\delta^{2}-\mu)^{2}},
		\end{align}
		which completes the proof.
		
		\section{Proof for Corollary~\ref{corollary}}\label{apex-e}
	
		According to~\eqref{decom}, $D_{\mathrm{KL}}(p_{o},p_{\boldsymbol{\theta}_{\mathcal{Z}}})$ is decomposed into the approximation and generalization errors. 
		Theorem~\ref{theorem1} demonstrates that there exits an optimized data-driven DL estimator $\mathbf{p}_{\boldsymbol{\theta}_{o}}(\mathbf{x})$ with
		at most $
		\lceil\log_{2}(d_{0}+1)\rceil
		$ hidden layers and sufficiently large $R$ such that the approximation error $
		J(p)-J(p_{\boldsymbol{\theta}_{o}})\leq\varepsilon$
		for any $\varepsilon>0$. 
		
		Moreover, Theorem~\ref{theorem2} implies that the generalization error $J(p_{\boldsymbol{\theta}_{o}})-J(p_{\boldsymbol{\theta}_{\mathcal{Z}}})$ satisfies 
		\begin{align}
			\lim_{|\mathcal{Z}|\rightarrow+\infty}\mathbf{P}\big([J(p_{\boldsymbol{\theta}_{o}})-J(p_{\boldsymbol{\theta}_{\mathcal{Z}}})]>\varepsilon\big)=0
		\end{align}
		for any $\varepsilon>0$. Combining Theorems~\ref{theorem1} and Theorem~\ref{theorem2}, we have
		\begin{align}
			\lim_{|\mathcal{Z}|\rightarrow+\infty}\mathbf{P}\big(D_{\mathrm{KL}}(p_{o},p_{\boldsymbol{\theta}_{\mathcal{Z}}})>\varepsilon\big)=0
		\end{align}
		for any $\varepsilon>0$, which completes the proof.
		\section{Proof for Theorem~\ref{theorem3}}\label{apex-f}
		Assume that $\Omega$ is fixed with $\sum_{m=1}^{|\Omega|}[J(\mathbf{f}_{\boldsymbol{\vartheta}}(\mathbf{x}_{m},\mathbf{H}_{m}))]^{2}\geq|\Omega|\delta_{u}$ for $\delta_{u}>0$, and then choose a collection of functions $\mathbf{f}_{1}(\mathbf{x},\mathbf{H}),\ldots,\mathbf{f}_{C_{u}}(\mathbf{x},\mathbf{H})\in\mathcal{D}_{u}$ such that $C_{u}=C_{u}(\varepsilon/16,\Theta_{u})$.
		According to Hoeffding's Inequality~\cite{mendenhall2012introduction} and Theorem~\ref{theorem2}, we have
		\begin{align}
			\mathbf{P}_{u}([J_{u}(\mathbf{f}_{\boldsymbol{\vartheta}_\Omega})-J_{u}(\mathbf{f}_{\boldsymbol{\vartheta}_o})]>\varepsilon
			|\,\Omega)&\leq\sum_{j=1}^{C_{u}}8\exp\bigg[-2(\frac{|\Omega|}{16}\varepsilon)^{2}/\sum_{m=1}^{|\Omega|}(2J(\mathbf{f}_{j}(\mathbf{x}_{m},\mathbf{H}_{m})))^{2}\bigg]\nonumber\\
			&\leq8\exp\bigg(\ln C_{u}-\frac{|\Omega|\varepsilon^{2}}{512\delta_{u}}\bigg).
		\end{align}
		Integrating out $\mathbf{P}_{u}([J_{u}(\mathbf{f}_{\boldsymbol{\vartheta}_\Omega})-J_{u}(\mathbf{f}_{\boldsymbol{\vartheta}_o})]>\varepsilon
		|\,\Omega)$ over $\Omega$ yields
		\begin{align}
			\mathbf{P}_{u}\big([J_{u}(\mathbf{f}_{\boldsymbol{\vartheta}_\Omega})-J_{u}(\mathbf{f}_{\boldsymbol{\vartheta}_o})]>\varepsilon
			\big)\leq8\exp\left(\ln C_{u}-\frac{|\Omega|\varepsilon^{2}}{512\delta_{u}}\right)+\mathbf{P}_{\Omega},
		\end{align}
		which completes the proof.
	\end{appendices}

	
	%%%%%%%%%%%%%%%%%%%%%%%%%%%%%%%%%%%%%%%%%%%%%%%%%%%%%%%%%%%%%%%%%%%%%%%%%%%%%%%
	%%%%%%%%%%%%%%%%%%%%%%%%%%%%%%%%%%%%%%%%%%%%%%%%%%%%%%%%%%%%%%%%%%%%%%%%%%%%%%%
	
	\linespread{1.083}
	\bibliography{inference}
	\bibliographystyle{IEEEtran}
	
\end{document}